\documentclass[prl,reprint, twocolumn,showpacs,preprintnumbers,amsmath,amssymb,nofootinbib,floatfix]{revtex4-1} %,tightenlines
\usepackage{graphicx}

\usepackage{mathrsfs}
\usepackage{hyperref}

\usepackage{slashed}

\usepackage{color}

\usepackage{gensymb}

\usepackage{amsmath}
\allowdisplaybreaks[4]

\def \matrix #1 {\left(\begin{array}{cc} #1 \end{array}\right)}

\def\II{\hbox{{1}\kern-.25em\hbox{l}}}

\begin{document}

\title{Next-to-Leading-Order Weak Annihilation Correction to Rare $B \to \left \{K, \pi \right \} \ell^{+} \ell^{-}$ Decays}

\author{Yong-Kang Huang$^{a}$}
\email{huangyongkang@mail.nankai.edu.cn}

\author{Yue-Long Shen$^{b}$}
\email{corresponding author: shenylmeteor@ouc.edu.cn}

\author{Chao Wang$^{c}$}
\email{corresponding author: chaowang@nankai.edu.cn}

\author{Yu-Ming Wang$^{a}$}
\email{corresponding author: wangyuming@nankai.edu.cn}

\affiliation{ ${}^a$ School of Physics, Nankai University,
Weijin Road 94, Tianjin 300071, P.R. China  \\
${}^b$ College of Information Science and Engineering, Ocean University of China,
Qingdao 266100, Shandong, China \\
${}^c$ Department of Mathematics and Physics,
Huaiyin Institute of Technology,
Meicheng East Road 1,  Huaian, Jiangsu 223200, P.R. China
}

\date{\today}

\begin{abstract}
\noindent

We accomplish for the first time the next-to-leading-order computation of the weak annihilation contribution
to the exclusive electroweak penguin decays  $B \to \left \{K, \pi \right \} \ell^{+} \ell^{-}$
with an energetic light-flavour meson, which is an essential missing piece of the complete QCD correction
to the matrix elements of hadronic operators in the weak effective Hamiltonian.
Both the hard functions and the jet functions  in the perturbative factorization formulae
from the short-distance fluctuations at the two distinct scales $m_b$ and $\sqrt{m_b \, \Lambda}$
are determined at ${\cal O}(\alpha_s)$ with the soft-collinear factorization technique.
We then demonstrate that the one-loop weak annihilation correction can bring about the noticeable impacts
on theory predictions for the CP asymmetries and the isospin asymmetry
in the $B \to \pi  \ell^{+}  \ell^{-}$ decays.
\\[0.4em]

%\noindent
%{\footnotesize{Keywords: Heavy quark decays, CP violation, Light-cone distribution amplitudes}}
\end{abstract}

%\preprint{TUM-HEP-1489/23}

\maketitle

%
%%%%%%%%%%%%%%%%%%%%%%%%%%%%%%%%%%%%%%%%%%%%%%%%%%%%%%%%%%%%%%%%%%%%%%%%%%%%
\section{Introduction}
%%%%%%%%%%%%%%%%%%%%%%%%%%%%%%%%%%%%%%%%%%%%%%%%%%%%%%%%%%%%%%%%%%%%%%%%%%%%
%

It is widely recognized that the exclusive electroweak penguin $B$-meson  decays are of extraordinary importance
to explore the intricate dynamics of the quark and lepton sectors in the Standard Model (SM)
and to probe sensitively the nonstandard effects with clean experimental signatures.
The prominent flavour anomalies observed in the flagship semileptonic $b \to s  \ell^{+} \ell^{-}$ decays
include the ${\cal O}(4 \, \sigma)$ deviations of the measured  ${\cal BR}(B^{+} \to K^{+} \mu^{+} \mu^{-})$
in the low-$q^2$ bins with respect to the SM predictions
and the ${\cal O}(2 \, \sigma)$ tensions between the LHCb measurements of the angular observable
$P_5^{\prime}(B^{0} \to K^{0 \ast} \mu^{+} \mu^{-})$ and the state-of-the-art SM computations
in the two anomalous  bins (see \cite{Capdevila:2023yhq} for an overview).
In addition, the first LHCb measurement of the integrated  isospin asymmetry $A_{\rm I}(B \to K  \mu^{+} \mu^{-})$
in the entire $q^2$-spectrum deviates from zero at the level of $4 \, \sigma$ \cite{LHCb:2012bin},
while the updated LHCb measurement of this asymmetry favours the lower significance of $1.5 \, \sigma$ \cite{LHCb:2014cxe},
due to the inclusion of the $2012$ data set and the improved analysis of the $2011$ data sample.
Precision measurements of the isospin asymmetries in the exclusive $b \to \left \{d , s \right \} \ell^{+} \ell^{-}$  decays
will be highly beneficial for constraining the Wilson coefficients of the $\Delta B=1$  four-quark operators
and for unraveling the ultimate nature of the spectator-scattering mechanism in  hard  exclusive reactions generally
\cite{Feldmann:2002iw,Khodjamirian:2012rm,Lyon:2013gba,Hou:2014dza,Hambrock:2015wka}
(see  \cite{Kagan:2001zk,Ball:2006eu} for discussions in the context of the  radiative $B$-meson decays).
As a consequence, the in-depth exploitation of the flavour structure of the weak effective theory
from the exclusive electroweak penguin $B$-meson decay constitutes one of the primary physics goals of
the LHCb and Belle II experiments, thus triggering the intense  theoretical activities  towards
controlling better  hadronic uncertainties of the SM predictions for more than one decade
(see  for instance \cite{Khodjamirian:2010vf,Gubernari:2020eft,Bobeth:2017vxj,Bharucha:2015bzk,
Gao:2019lta,Descotes-Genon:2023ukb,Lu:2018cfc,Cui:2022zwm,Huang:2023jdu,Khodjamirian:2023wol}).

The field-theoretical framework to systematically address the hadronic matrix elements of the electroweak penguin $B$-meson decays
at large recoil has been constructed in the heavy quark mass limit with the perturbative factorization  formalism \cite{Beneke:2001at,Beneke:2004dp}
(see also \cite{Becher:2005fg,Ali:2006ew} for an equivalent formulation in soft-collinear effective theory (SCET)).
The resulting factorization formulae of the exclusive $b \to \left \{d , s \right \} \ell^{+} \ell^{-}$  decay amplitudes
can be expressed in terms of  heavy-to-light form factors,  light-cone distribution amplitudes (LCDAs),
and perturbatively calculable short-distance coefficient functions,
in analogy to the factorized expressions for the charmless hadronic $B$-meson decays \cite{Beneke:1999br,Beneke:2000ry}.
In particular, both the weak annihilation  and  spectator-scattering  topologies  due to the hard-collinear photon
radiation off the light spectator quark yield vanishing contributions to the transverse amplitude of $B \to V \ell^{+} \ell^{-}$
at leading power in the $\Lambda/m_b$ expansion \cite{Grinstein:2000pc,Beneke:2001at}.
This interesting observation motivates further  factorization analysis of the particular subleading power corrections
to the transverse amplitude dependent of the electric charge of the spectator quark \cite{Kagan:2001zk,Feldmann:2002iw,Beneke:2004dp},
which are fortunately determined by  the soft-collinear convolution integrals without the disturbing end-pointed divergences
(with the exception of the matrix element of the chromomagnetic dipole operator).
However,   the factorization-scale invariance  of both the transverse and longitudinal amplitudes of $B \to V \ell^{+} \ell^{-}$ was
not completely achieved at the next-to-leading order (NLO) in QCD even in the leading-power approximation,
owing to the unknown two-loop matrix elements of QCD penguin operators and  the absence of the one-loop
weak annihilation contribution in the longitudinal amplitude.
Moreover, extending the QCD computation of  the leading-power weak annihilation effect in
$B \to \left \{ V_{\|}, P  \right \} \ell^{+} \ell^{-}$ to the NLO accuracy
is in high demand for unraveling factorization properties of the exclusive heavy-hadron decay amplitudes
and for deepening our understanding towards model-independent properties of the $B$-meson distribution amplitudes.

In order to probe the weak annihilation mechanism of the  electroweak penguin bottom-meson decays
in a theoretically clean fashion, we will concentrate on the rigorous NLO QCD computation of the weak annihilation contribution
to $B \to \left \{K, \pi \right \} \ell^{+} \ell^{-}$ at leading power in the heavy quark expansion in this Letter,
by employing the modern SCET technique.
It is apparent that  such higher-order corrections to the exclusive $B \to \pi \ell^{+} \ell^{-}$ decay amplitudes
become more pronounced than the counterpart $B \to K \ell^{+} \ell^{-}$  decays,
thanks to the  unsuppressed contributions from the  current-current operators in the weak effective Hamiltonian.
Subsequently, we  report on a distinctive feature of the obtained soft-collinear factorization formula at ${\cal O}(\alpha_s)$,
in contrast with the analytical structure of the available leading-order (LO) expression.
Phenomenological implications of the newly computed NLO weak annihilation contributions
to the semileptonic $B \to \left \{K, \pi \right \} \ell^{+} \ell^{-}$ decay observables
will be further explored with the improved lattice determinations of the leading-twist pion and kaon LCDAs
as well as one sample model for the two-particle $B$-meson LCDAs in heavy quark effective theory (HQET).

%%%%%%%%%%%%%%%%%%%%%%%%%%%%%%%%%%%%%%%%%%%%%%%%%%%%%%%%%%%%%%%%%%%%%%%%%%%%
\section{General analysis}
%%%%%%%%%%%%%%%%%%%%%%%%%%%%%%%%%%%%%%%%%%%%%%%%%%%%%%%%%%%%%%%%%%%%%%%%%%%%

The effective weak Hamiltonian for  the  semileptonic $b \to D  \ell^{+}  \ell^{-}$ (with $D=d, \, s$) transitions
in the SM can be written as \cite{Beneke:2004dp}
\begin{eqnarray}
{\cal H}_{\rm eff} &=&  - {G_F \over \sqrt{2}} \, \left [ V_{t b} V_{t D}^{\ast} \,  {\cal H}_{\rm eff}^{(t)}
+ V_{u b} V_{u D}^{\ast} \,  {\cal H}_{\rm eff}^{(u)} \right ] + {\rm h.c.}   \,,
\label{effective weak Hamiltonian of b to D ll}
\end{eqnarray}
by employing  the unitarity relations of the Cabibbo-Kobayashi-Maskawa (CKM) matrix elements,
and
\begin{eqnarray}
{\cal H}_{\rm eff}^{(t)} &=& C_1 \, {\cal Q}_1^{c} + C_2 \, {\cal Q}_2^{c} + \sum_{i=3,..., 10} C_i \, {\cal Q}_i \,,
\nonumber \\
{\cal H}_{\rm eff}^{(u)} &=& C_1 \, ({\cal Q}_1^{c} - {\cal Q}_1^{u})  + C_2 \, ({\cal Q}_2^{c} - {\cal Q}_2^{u})  \,.
\end{eqnarray}
%We  will adopt the operator basis proposed in \cite{Chetyrkin:1997gb} for the sake of implementing
%the naive dimensional regularization  scheme of $\gamma_5$ consistently.
The hadronic matrix elements of the two semileptonic operators ${\cal Q}_{9, 10}$ can be readily expressed in terms of
the heavy-to-light $B$-meson decay form factors \cite{Beneke:2000wa}.
The remaining contribution to the electroweak penguin decay amplitude
can be determined  by the contraction of  the exclusive $B \to  \left \{K, \pi \right \}  \, \gamma^{\ast}$  matrix element
with the vector lepton current $\gamma^{\ast} \to \ell^{+} \ell^{-}$.
We are then led to introduce
\begin{eqnarray}
\langle P(p^{\prime}) \, \gamma^{\ast}(q, \mu) | {\cal H}_{\rm eff}^{(t, \, u)}  |  \bar B(p) \rangle
= - {g_{\rm em} \, m_b  \over 4 \, \pi^2} \, { \mathcal{T}_P^{(t, \, u)}(q^2)  \over m_B} &&
\nonumber \\
\times \, \left [q^2 \, (p_{\mu} + p^{\prime}_{\mu}) - (m_B^2 - m_P^2)  \, q_{\mu} \right ], &&
\hspace{0.5 cm}
\end{eqnarray}
where $| P \rangle$ represents $| K^{-} \rangle$ ($| \pi^{-} \rangle$) for the charged $B^{-}$ decay,
and $| \bar K^{0} \rangle$ ($-\sqrt{2} \, \pi^{0}$) for the neutral $\bar B_{d}^{0}$ decay.
The resulting factorization formulae for the hadronic quantities $\mathcal{T}_P^{(t, \, u)}$ at leading power
can be cast in the form %\cite{Beneke:2001at,Beneke:2004dp}
%\begin{widetext}
\begin{eqnarray}
&&  \mathcal{T}_P^{(t, \, u)} = C_P^{(t, \, u)}(q^2) \, f_{B P}^{+}(q^2)
- {\pi^2 \over N_c} \, {\mathcal{F}_B \, f_P \over m_B} \,
\sum_{m= \pm}
\nonumber \\
&&  \int_0^{\infty} {d \omega \over \omega}  \,  \int_0^1 d u \,
T_{P, \, m}^{(t, \, u)}(\omega, u, \mu) \,  \phi_{B, m}(\omega, \mu) \,  \phi_P(u, \mu) \,,
\hspace{0.3 cm}
\label{QCDF for non-form-factor effects}
\end{eqnarray}
%\end{widetext}
where $f_{B P}^{+}(q^2)$ denotes the vector form factor  defined by the local QCD matrix element
$\langle P(p^{\prime}) |  \bar D  \gamma_{\mu} b |  \bar B(p) \rangle$.
$\mathcal{F}_B$ corresponds to  the HQET $B$-meson decay constant and $f_P$ is the scale-independent decay constant
of the energetic pseudoscalar meson.
Furthermore $\phi_{B, \pm}(\omega, \mu)$ and $\phi_P(u, \mu)$ refer to the  $B$-meson and light-meson LCDAs \cite{Grozin:1996pq,Lepage:1979zb,Efremov:1979qk,Lepage:1980fj}.
The short-distance coefficient functions $C_P^{(i)}(q^2)$  and $T_{P, \,  m}^{(i)}(\omega, u, \mu)$ $(i =u, \, t)$
can be  expanded perturbatively  as follows
(similarly for any other partonic quantity)
\begin{eqnarray}
C_P^{(i)} = \sum_{\ell}\left ( {\alpha_s \over 4 \pi} \right )^{\ell}   \, C_P^{(\ell,i)}\,,
\hspace{0.3 cm}
T_{P, m}^{(i)} =    \sum_{\ell}\left ( {\alpha_s \over 4 \pi} \right )^{\ell}   \, T_{P, m}^{(\ell,i)} \,.
\hspace{0.5 cm}
\end{eqnarray}
The analytical expressions of the known perturbative kernels at  ${\cal O}(\alpha_s)$
can be extracted from \cite{Beneke:2001at,Beneke:2004dp}.
In particular, only the subleading twist HQET distribution amplitude $\phi_{B, -}(\omega, \mu)$
appears in the tree-level factorization formula for the spectator-scattering contribution of $B \to P \ell^{+} \ell^{-}$,
in sharp contrast with the radiative leponic  decays $B \to \gamma \ell \bar \nu_{\ell}$,
$B \to \gamma \ell^{+} \ell^{-}$ \cite{Descotes-Genon:2002crx,Descotes-Genon:2002lal,Bosch:2003fc,Beneke:2020fot},
and the double radiative  decay $B_{d, s} \to \gamma \gamma$ \cite{Descotes-Genon:2002lal,Bosch:2002bv,Shen:2020hfq,Qin:2022rlk}.
As discussed in \cite{Beneke:2001at,Beneke:2004dp}, it turns out to be advantageous to introduce two effective Wilson coefficients
independent of the renormalization conventions for the weak effective Hamiltonian
\begin{eqnarray}
{\cal C}_{9, P}^{(i)}(q^2) = C_9 \, \delta^{i t}+ {2 m_b \over m_B} \, \frac{\mathcal{T}_P^{(i)}(q^2)}{ f_{B P}^{+}(q^2)} \,.
%\nonumber \\
%{\cal C}_{9, P}^{(u)}(q^2) = {2 m_b \over m_B} \, \frac{T_P^{(t)}(q^2)}{ f_{B P}^{+}(q^2)} \,.
\end{eqnarray}

%%%%%%%%%%%%%%%%%%%%%%%%%%%%%%%%%%%%%%%%%%%%%%%%%%%%%%%%%%%%%%%%%%%%%%%%%%%%
\section{QCD factorization for the weak annihilation contribution}
%%%%%%%%%%%%%%%%%%%%%%%%%%%%%%%%%%%%%%%%%%%%%%%%%%%%%%%%%%%%%%%%%%%%%%%%%%%%

%%%%%%%%%%%%%%%%%%%%%%%%%%%%%%%%%%%%%%%%%%%%%%%%%%%%%%%%%%%%%%%
\begin{figure}[tp]
\begin{center}
\includegraphics[width=0.95 \columnwidth]{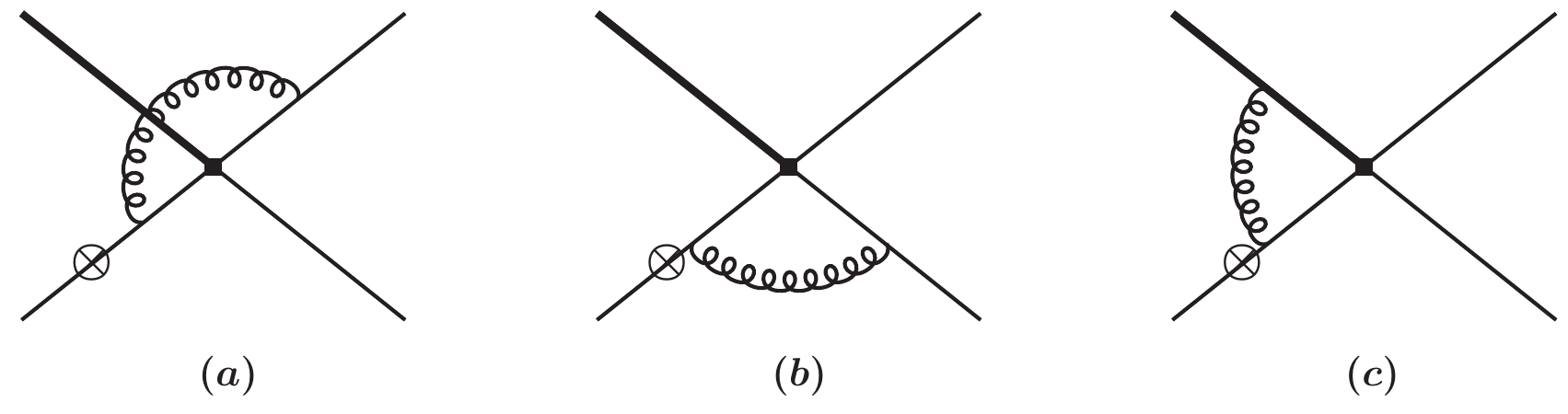}
\vspace*{0.1cm}
\caption{Sample Feynman diagrams for the NLO weak annihilation contribution to $B \to \left \{K, \pi \right \} \ell^{+} \ell^{-}$,
where the circled cross indicates the hard-collinear photon radiation.
Note that only the virtual photon emission from the soft spectator quark yields the leading-power contribution
in the $\Lambda/m_b$ expansion. }
\label{fig: NLO weak annihilation diagrams for B to P ll}
\end{center}
\end{figure}
%%%%%%%%%%%%%%%%%%%%%%%%%%%%%%%%%%%%%%%%%%%%%%%%%%%%%%%%%%%%%%%

We are now in a position to compute the NLO weak annihilation contribution to $B \to \left \{K, \pi \right \} \ell^{+} \ell^{-}$
with the soft-collinear factorization technique.
Inspecting the one-loop partonic diagrams displayed in Figure \ref{fig: NLO weak annihilation diagrams for B to P ll}
indicates that the four distinct classes of  field modes (hard, hard-collinear,  anti-hard-collinear, and soft)
carrying the four-momenta   $P_{{\rm h}, \, \mu} \sim {\cal O}(1, 1, 1)$,  $P_{{\rm hc}, \, \mu} \sim {\cal O}(1, \lambda, \lambda^{1/2})$,
$P_{{\rm  \overline{hc}}, \, \mu} \sim {\cal O}(\lambda, 1, \lambda^{1/2})$
and $P_{{\rm s}, \, \mu} \sim {\cal O}(\lambda, \lambda, \lambda)$ appear in the problem under consideration,
where the individual momentum components correspond to $n \cdot P$, $\bar n \cdot P$ and $P_{\perp}$ for an arbitrary momentum $P_{\mu}$.
The two light-cone vectors $n_{\mu}$ and $\bar n_{\mu}$ satisfying $n^2=\bar n^2=0$ and $n \cdot \bar n$=2
are chosen such that $n \cdot q \sim \bar n \cdot p^{\prime} \sim {\cal O}(m_b)$.
Additionally,  the heavy-quark expansion parameter  $\lambda$ scales as $\Lambda_{\rm QCD} / m_b$
in our power-counting analysis. %\cite{Beneke:2002ph}.

A systematic decoupling of the hard and (anti)-hard-collinear fluctuations can be  achieved  by taking advantage of
the two-step matching strategy ${\rm QCD} \to {\rm SCET_{I}} \to {\rm SCET_{II}}$.
In accordance with the power-counting argument developed in the position-space formalism \cite{Beneke:2003pa,Beneke:2005vv,Beneke:2006mk},
both the ${\rm A0}$-type and ${\rm B1}$-type ${\rm SCET_{I}}$ operators can generate the leading-power contribution
to the weak annihilation amplitude.
It is then straightforward to write down the ${\rm SCET_{I}}$ representation of
the current-current and QCD penguin operators
${\cal Q}_i \in \left \{ {\cal Q}_{1}^{u},   {\cal Q}_{2}^{u}, {\cal Q}_{3-6}  \right \}$
\begin{eqnarray}
{\cal Q}_i = H_i^{\rm I} \star {\cal O}^{(\rm A0)}  + H_i^{\rm II}  \star {\cal O}^{(\rm B1)}  \,,
\end{eqnarray}
for the ``right-insertion" topology where the spinor indices are contracted along the quark lines of
the bottom meson and the light meson.
We have employed here the  customary asterisk notation in SCET
\begin{eqnarray}
H_i^{\rm I} \star {\cal O}^{(\rm A0)}
&=& \int d \hat{t} \, H_i^{\rm I}(\hat{t}) \, {\cal O}^{(\rm A0)}(t)  \,,
\nonumber \\
H_i^{\rm II}  \star {\cal O}^{(\rm B1)}
&=&  \int d \hat{t} \,  d \hat{s} \, H_i^{\rm II}(\hat{t},   \hat{s})  \, {\cal O}^{(\rm B1)}(t,   s) \,,
\end{eqnarray}
with the dimensionless convolution variables $\hat{t} = (\bar n \cdot p^{\prime}) \, t$
and $\hat{s} = (n \cdot q) \, s$.
The explicit expressions of the two ${\rm SCET_{I}}$ operators of our interest are given by
\begin{eqnarray}
{\cal O}^{(\rm A0)}  &=&  \left [ \left (\bar \chi W_{\bar c} \right ) (t \bar n)  \,  {\slashed{\bar n} \over 2} \,
(1- \gamma_5)  \,  \left (W_{\bar c}^{\dagger} \chi \right ) (0)  \right ]
\nonumber \\
&& \left [ \left (  \bar \xi W_{c} \right )(0) \, \slashed{ n}  \, (1 - \gamma_5) \, h_v(0) \right ],
\\
{\cal O}^{(\rm B1)} &=& {1 \over m_b} \, \left [ \left (\bar \chi W_{\bar c} \right ) (t \bar n)  \, {\slashed{\bar n} \over 2} \,
(1- \gamma_5) \, \left (W_{\bar c}^{\dagger} \chi \right ) (0)  \right ]
\nonumber \\
&& \left [ \left (  \bar \xi W_{c} \right )(0) \, {\slashed n \over 2}   \,
\left [ W_{c}^{\dagger}  i \slashed{D}_{\perp c}   W_{c} \right ](s n) \, (1 + \gamma_5) \, h_v(0) \right ],
\nonumber
\end{eqnarray}
where the collinear Wilson lines $W_{c}$ and  $W_{\bar c}$
are introduced to ensure gauge invariance \cite{Beneke:2003pa}.
Along the same vein, the matching equation for the ``wrong-insertion" topology  takes the form
\begin{eqnarray}
{\cal Q}_i = \tilde{H}_i^{\rm I} \star {\cal O}^{(\rm A0)}  + \tilde{H}_i^{\rm II}  \star {\cal O}^{(\rm B1)}  \,,
\end{eqnarray}
where both the bottom and light mesons  are  made up of the quark-antiquark pair from two different bilinears in the QCD operators
(see \cite{Beneke:2005vv}  for further discussions).
Performing the Fourier transformation with respect to the light-cone variables $\hat{t}$ and $\hat{s}$
defines the momentum-space short-distance functions
\begin{eqnarray}
\left \{ \mathbb{H}_i^{\rm I},  \, \mathbb{\tilde{H}}_i^{\rm I} \right \}(u)
&=& \int d \hat{t}  \, e^{i u \hat{t}} \, \left \{  H_i^{\rm I},  \, \tilde{H}_i^{\rm I} \right \} (\hat{t})\,,
\nonumber   \\
\left \{ \mathbb{H}_i^{\rm II},  \, \mathbb{\tilde{H}}_i^{\rm II} \right \}(u, \tau)
&=& \int d \hat{t}  \, e^{i (u \, \hat{t} + \tau \, \hat{s})} \,
\left \{  H_i^{\rm II},  \, \tilde{H}_i^{\rm II}\right \}(\hat{t}, \hat{s}) \,.
\hspace{0.8 cm}
\end{eqnarray}
It becomes apparent that we are required to derive both the ``right-insertion" and the ``wrong-insertion"
matching coefficients for the current-current operators ${\cal Q}_{1, 2}^{u}$,
but merely the ``wrong-insertion" kernels for the QCD penguin operators ${\cal Q}_{3-6}$.
Applying the technical framework of determining  the renormalized hard-scattering kernels
with the inclusion of the evanescent operators  detailed in \cite{Beneke:2009ek,Bell:2020qus}
leads to the desired results at the ${\cal O} (\alpha_s)$ accuracy
\begin{eqnarray}
\mathbb{H}_1^{\rm I(0)} &=&  0,   \hspace{0.4 cm}  \mathbb{H}_2^{\rm I(0)} = 1,
\hspace{0.4 cm} \mathbb{\tilde{H}}_{1, 4}^{\rm I(0)} =  {C_F \over N_c},
\hspace{0.4 cm}  \mathbb{\tilde{H}}_{2, \, 3}^{\rm I(0)} =  {1 \over N_c},
\nonumber  \\
\mathbb{\tilde{H}}_5^{\rm I(0)}  &=& {16 \over N_c},   \hspace{3.2 cm}  \mathbb{\tilde{H}}_6^{\rm I(0)}  =  16 \, {C_F \over N_c},
\nonumber  \\
\mathbb{H}_1^{\rm II(0)} &=&  {1 \over N_c} {1 \over \bar u},
\hspace{3.0 cm}
\mathbb{H}_2^{\rm II(0)} =  2,
\\
\mathbb{\tilde{H}}_{1, 4}^{\rm II(0)} &=&  {1 \over N_c} \, \left [ 2 \, C_F -  {1 \over N_c} {1 \over \bar u} \right ],
\qquad
\mathbb{\tilde{H}}_{2, 3}^{\rm II(0)} = {2 \over N_c} \, \left [ 1 +  {1 \over \bar u} \right ],
\nonumber \\
\mathbb{\tilde{H}}_5^{\rm II(0)} &=&  16 \, \mathbb{\tilde{H}}_{2}^{\rm II(0)},
\hspace{2.5 cm}
\mathbb{\tilde{H}}_6^{\rm II(0)} =  16 \, \mathbb{\tilde{H}}_{1}^{\rm II(0)},
\nonumber
\end{eqnarray}
and
\begin{eqnarray}
\mathbb{H}_1^{\rm I(1)} &=&  {C_F \over 2  N_c} \, \left [ -6 \, \mathcal{L}  + \mathfrak{h}(u) \right ],
\nonumber \\
\mathbb{H}_2^{\rm I(1)} &=&  -{C_F \over 2} \, \left [ \mathcal{L}^2  +  5 \, \mathcal{L} + {\pi^2 \over 6} + 12 \right ],
\nonumber \\
\mathbb{\tilde{H}}_1^{\rm I(1)} &=&   {1 \over N_c}  \left [C_F  \mathbb{H}_2^{\rm I(1)} -   \mathbb{H}_1^{\rm I(1)} \right ] - C_F,
\nonumber \\
\mathbb{\tilde{H}}_2^{\rm I(1)} &=& \mathbb{\tilde{H}}_3^{\rm I(1)}
= {1 \over N_c} \,  \mathbb{H}_2^{\rm I(1)} + 2 \, \mathbb{H}_1^{\rm I(1)},
 \\
\mathbb{\tilde{H}}_4^{\rm I(1)} &=&  {1 \over N_c}   \left [C_F  \mathbb{H}_2^{\rm I(1)} -   \mathbb{H}_1^{\rm I(1)} \right ]
+  {C_F \over N_c} \, \left [ {2 \over 3} \, n_f - N_c  \right ],
\nonumber \\
\mathbb{\tilde{H}}_5^{\rm I(1)} &=&  {16 \over N_c} \,  \mathbb{H}_2^{\rm I(1)} +  32\,  \mathbb{H}_1^{\rm I(1)},
\nonumber \\
\mathbb{\tilde{H}}_6^{\rm I(1)} &=&  {16 \over N_c}  \left [C_F  \mathbb{H}_2^{\rm I(1)} -   \mathbb{H}_1^{\rm I(1)} \right ]
+  {10 \, C_F \over N_c}  \left [ {2 \over 3}  n_f - 4 N_c  \right ].
\nonumber
\end{eqnarray}
For brevity we have introduced the following conventions
\begin{eqnarray}
\mathfrak{h}(u)  &=& \ln^2  \left (u \bar u \right )  + \left (3 - {2 \over u} \right ) \, \ln \bar u
 - 2 \, \ln^2 u + \left (3 - {1 \over \bar u} \right ) \, \ln  u
\nonumber \\
&& + \,  4  \, {\rm Li}_{2}(u) - {\pi^2 \over 3} - 22  +  i  \,  \pi \,  \left ( 2 \,  \ln \bar u - 2  \, \ln u  - 3 \right ),
\nonumber \\
\mathcal{L} &\equiv&  \ln \left ( {\mu^2 \over m_b^2} \right ) \,, \qquad \bar u \equiv 1-u  \,,
\qquad n_f = 5.
\end{eqnarray}

As demonstrated  in \cite{Chay:2003ju,Bauer:2004tj,Beneke:2005vv}, the hard-collinear and aniti-hard-collinear fields decouple
already at the hard scale of ${\cal O} (m_b)$.
This essential observation can be attributed to the disappearing leading-power interactions between
the (hard)-collinear fields with different directions in the SCET Lagrangian after the field redefinition \cite{Bauer:2001yt}
(see also \cite{Becher:2014oda,Beneke:2015wfa} for an overview).
The obtained ${\rm SCET_{I}}$ matrix elements of  ${\cal O}^{(\rm A0)}$ and ${\cal O}^{(\rm B1)}$
then fall apart into two distinct sectors each, thus generating the non-local effective matrix elements below
\begin{widetext}
\begin{eqnarray}
&& \langle P(p^{\prime})| \left (\bar \chi W_{\bar c} \right ) (t \bar n)  \,  {\slashed{\bar n} \over 2} \,
(1- \gamma_5)  \,  \left (W_{\bar c}^{\dagger} \chi \right ) (0) | 0 \rangle
 = i \, f_P \,  {\bar n \cdot p^{\prime} \over 2} \,
\int_0^1 d u \, e^{i  \, (\bar n \cdot p^{\prime}) \, u \, t} \,  \phi_P(u, \mu),
\nonumber \\
&& \int d^4 x \, e^{i q \cdot x }  \langle 0 | {\rm T} \left \{ j_{\rm em, \, \mu}^{\|}(x),
\left (  \bar \xi W_{c} \right )(0)   \gamma_5 h_v(0) \right \} |  \bar B(p) \rangle
 = {\cal J}_{P}^{(\rm A0)} (n \cdot q, \bar n \cdot q) \, \bar n_{\mu},
\\
&& {n \cdot q \over 2 \, \pi}  \int d^4 x \, e^{i q \cdot x }  \int d s \, e^{i (n \cdot q)  \tau  s} \,
\langle 0 | {\rm T} \left  \{ j_{\rm em, \, \mu}^{\|}(x),
\left (  \bar \xi W_{c} \right )(0)   \gamma_5
\left [ W_{c}^{\dagger}  i \slashed{D}_{\perp c}   W_{c} \right ](s n)  \, h_v(0) \right  \} |  \bar B(p) \rangle
= {\cal J}_{P}^{(\rm B1)} (n \cdot q, \bar n \cdot q, \tau) \, \bar n_{\mu}.
\nonumber
\end{eqnarray}
\end{widetext}
The ${\rm SCET_{I}}$ representation of the electromagnetic current relevant here is given by
\begin{eqnarray}
j_{\rm em, \,  \mu}^{\|} &\supset&  j_{\rm em, \,  \mu}^{\|, \, (0)}  + j_{\rm em, \,  \mu}^{\|, \, (2)}
 \\
&=& \bar \xi  {\slashed {n} \over 2} \xi  \,  \bar n_{\mu}
+ \left [  (\bar \xi  W_c)   {\slashed {n} \over 2}   (Y_s^{\dagger} q_s)
+ (\bar q_s  Y_s)   {\slashed {n} \over 2}  (W_c^{\dagger} \xi) \right ]  \bar n_{\mu},
\nonumber
\end{eqnarray}
where the second term on the right-hand side converts the soft spectator quark into a hard-collinear quark
and counts as ${\cal O}(\lambda^2)$.
The leading-power current $j_{\rm em, \,  \mu}^{\|, \, (0)}$ can contribute to ${\cal J}_{P}^{(\rm A0)}$ (${\cal J}_{P}^{(\rm B1)} $)
only through time-ordered products with an additional SCET Lagrangian ${\cal L}_{\xi q_s}^{(2)}$ (${\cal L}_{\xi q_s}^{(1)}$) \cite{Beneke:2002ni}.
By contrast,  the subleading-power current $ j_{\rm em, \,  \mu}^{\|, \, (2)}$ can bring about the non-vanishing
contribution to ${\cal J}_{P}^{(\rm A0)}$ already at tree level (and certainly at higher orders in $\alpha_s$ with insertions of
the unsuppressed SCET Lagrangian ${\cal L}_{\xi}^{(0)}$).
%$Y_s (x) = {\rm P} \, {\rm exp} \left [ i \, g_s  \int_{-\infty}^{0} d s \, \bar n \cdot A_s(x + s \bar n)\right ]$.
Integrating out the hard-collinear fluctuations in the ${\rm SCET_{I}}$ correlation functions
by virtue of the second-step matching ${\rm SCET_{I}} \to {\rm SCET_{II}}$
enables us to derive the soft-collinear factorization formulae \cite{Gao:2019lta,Wang:2021yrr}
\begin{eqnarray}
{\cal J}_{P}^{(\rm A0)} &=& \frac{\mathcal{F}_B  m_B}{2} \int_0^{\infty} \, d \omega \,
{\mathbb{J}_{\|, \, -}(n \cdot q, \bar n \cdot q, \omega)  \over \bar n \cdot q - \omega + i \, 0}  \phi_{B, -}(\omega, \mu),
\nonumber \\
{\cal J}_{P}^{(\rm B1)} &=& \frac{\mathcal{F}_B  m_B}{2}  \int_0^{\infty} \, d \omega \,
\mathbb{J}_{\|, \, +}(n \cdot q, \bar n \cdot q, \omega, \tau)  \phi_{B, +}(\omega, \mu).
\nonumber \\
\label{soft-collinear factorization formulae}
\end{eqnarray}
Employing the SCET Feynman rules collected in \cite{Beneke:2018rbh},
we can readily determine the renormalized jet functions $\mathbb{J}_{\|, \, -}$  and
$\mathbb{J}_{\|, \, +}$  up to the  order $\alpha_s$
\begin{eqnarray}
&& \left \{ \mathbb{J}_{\|, \, -}^{(0)}, \, \mathbb{J}_{\|, \, +}^{(0)} \right \}  =
\left \{ 1, \,  0  \right \},
\nonumber \\
&& \mathbb{J}_{\|, \, -}^{(1)} = C_F \,  \bigg  [ \ln^2    {\hat{\mu}^2  \over   \omega - \bar n \cdot q }
- 2 \,  \ln   {\hat{\mu}^2  \over   \omega - \bar n \cdot q }   \,  \ln  (1 + \eta)
\nonumber \\
&&  \hspace{1.0 cm}
-  \ln^2  (1 + \eta) +  \left ( { 2 \over \eta} - 1 \right )  \,  \ln  (1 + \eta) -  {\pi^2 \over 6} - 1  \bigg  ],
\nonumber \\
&& \mathbb{J}_{\|, \, + }^{(1)} =  2 \,  C_F \, {n \cdot q \over \omega} \,
\ln  (1 + \eta)  \, (1-\tau) \, \theta(\tau) \, \theta(1-\tau),
\end{eqnarray}
where we have defined two variables $\hat{\mu}^2 = \mu^2/ n \cdot q$ and $\eta=- \omega / \bar n \cdot q$.
Interestingly, the leading-twist bottom-meson distribution amplitude $\phi_{B, +}(\omega, \mu)$ now enters
the perturbative factorization formula (\ref{soft-collinear factorization formulae})
of the weak annihilation amplitude  at ${\cal O}(\alpha_s)$,
in contrast with the factorized expression for the tree-level annihilation amplitude.
On the basis of the asymptotic behaviours of $\phi_{B, \pm}(\omega, \mu)$ \cite{Grozin:1996pq,Lange:2003ff,Braun:2003wx,Bell:2013tfa,Lee:2005gza,Feldmann:2014ika,Braun:2017liq,Feldmann:2023aml},
we can further verify that the soft-collinear convolution integrals in (\ref{soft-collinear factorization formulae}) converge
at $q^2 \sim {\cal O}(m_b \, \Lambda_{\rm QCD})$ and the complex-valued invariant functions ${\cal J}_{P}^{(\rm A0)}$
and ${\cal J}_{P}^{(\rm B1)}$ cannot be entirely described by the two particular moments $\lambda_{B, \, +}^{-1}$
and $\lambda_{B, \, -}^{-1}(q^2)$ of the bottom-meson distribution amplitudes  as previously introduced in \cite{Beneke:2001at}.

We now summarize the explicit expressions of the NLO weak annihilation corrections
to the short-distance matching functions $T_{P, \,  m}^{(t, \, u)}$ appearing
in the QCD factorization formulae (\ref{QCDF for non-form-factor effects})
\begin{eqnarray}
\left \{ T_{P,   +}^{(u)},  \,\,  T_{P,  +}^{(t)} \right \}
& \supset &   - Q_q  N_c    {2 m_B \over m_b}
\int_0^1 d \tau
\mathbb{J}_{\|, +}(n \cdot q, \bar n \cdot q, \omega, \tau)
\nonumber \\
&& \hspace{-0.8  cm}
\bigg  \{- \sum_{i=1}^{2}   C_i  \left [  \mathbb{H}_i^{\rm II}(u, \tau) \, \delta_{q u}
- \mathbb{\tilde{H}}_i^{\rm II}(u, \tau) \, \delta_{q d}  \, \delta_{P \pi}\right ],
\nonumber \\
&& \hspace{-0.5  cm}
\sum_{i=3}^{6}   C_i \, \mathbb{\tilde{H}}_i^{\rm II}(u, \tau)   \bigg  \},
\nonumber \\
\left \{ T_{P,  -}^{(u)},  \,\,   T_{P,   -}^{(t)} \right \}
& \supset &   Q_q  N_c   {4 m_B \over m_b}
{\mathbb{J}_{\|,  -}(n \cdot q, \bar n \cdot q, \omega)    \over \hat{q}^2  - 1 + i \, 0}
\nonumber \\
&&  \hspace{-0.8 cm}
\bigg  \{ - \sum_{i=1}^{2}  \, C_i \, \left [  \mathbb{H}_i^{\rm I}(u) \, \delta_{q u}
-   \mathbb{\tilde{H}}_i^{\rm I}(u) \, \delta_{q d} \, \delta_{P \pi} \right ],
\nonumber \\
&& \hspace{-0.5  cm}
\sum_{i=3}^{6} \,  C_i \, \mathbb{\tilde{H}}_i^{\rm I}(u)  \bigg \},
\label {Final results: NLO Weak Annihilation Effects}
\end{eqnarray}
where $Q_q$ denotes the electric charge of the soft spectator quark in the $B$-meson.
Furthermore, $\hat{q}^2=q^2 /(m_B \, \omega)$,
$\delta_{P \pi}=1, \, 0$ for $\pi$ and $K$ mesons, respectively.
We remark in passing that the hard coefficient functions $\mathbb{H}_{1, 2}^{\rm II}$
and $\mathbb{\tilde{H}}_{1-6}^{\rm II}$ of the ${\rm B1}$-type ${\rm SCET_{I}}$ operators
are independent of the variable $\tau$  at ${\cal O}(\alpha_s^0)$
and the jet function $\mathbb{J}_{\|, +}$ starts from ${\cal O}(\alpha_s)$,
thus allowing us to perform the integration over $\tau$ in  the obtained  $ T_{P,   +}^{(u, \, t)}$
straightforwardly  in the NLO approximation.

%%%%%%%%%%%%%%%%%%%%%%%%%%%%%%%%%%%%%%%%%%%%%%%%%%%%%%%%%%%%%%%%%%%%%%%%%%%%
\section{Numerical  analysis}
%%%%%%%%%%%%%%%%%%%%%%%%%%%%%%%%%%%%%%%%%%%%%%%%%%%%%%%%%%%%%%%%%%%%%%%%%%%%

We continue to explore phenomenological implications of the NLO weak annihilation contributions
to the exclusive rare decays $B \to \left \{K, \pi \right \} \ell^{+} \ell^{-}$ at large hadronic recoil.
The Wilson coefficients $C_i$ in the weak effective Hamiltonian (\ref{effective weak Hamiltonian of b to D ll})
will be evaluated at the initial scale $\mu_{W} = m_W$ \cite{Chetyrkin:1996vx,Bobeth:1999mk}
and then evolved to the hard scale $\mu_{\rm h} = m_b$ with the required accuracy (following \cite{Bobeth:2003at,Huber:2005ig}).
We further adopt  the three-parameter ans\"{a}tz  for the HQET $B$-meson LCDAs as  proposed in \cite{Beneke:2018wjp}
(see \cite{Galda:2022dhp,Feldmann:2022uok,Braun:2018fiz,Wang:2019msf} for additional discussions),
which can be analytically evolved into the hard-collinear scale with the renormalization-group (RG) formalism
at the leading-logarithmic accuracy.
The leptonic decay constants of the $B$-meson, pion and kaon are taken from the averages of the lattice QCD simulations \cite{FlavourLatticeAveragingGroupFLAG:2021npn}.
Moreover, we will  truncate  the Gegenbauer expansion of the leading-twist pion and kaon LCDAs
by including only the  two lowest moments whose intervals have been determined from lattice QCD
at the $\overline{\rm MS}$ scale $\mu_0=2.0 \, {\rm GeV}$ with full control of all systematic uncertainties \cite{RQCD:2019osh}.
The two-loop RG evolution of these Gegenbauer moments \cite{Mueller:1993hg,Mueller:1994cn} (see also \cite{Agaev:2010aq,Wang:2017ijn})
will be then implemented for the sake of describing their scale dependence.
In particular, we employ the improved QCD predictions of the semileptonic form factors $f_{B \pi}^{+}(q^2)$ and $f_{B K}^{+}(q^2)$
displayed in \cite{Cui:2022zwm}.
The choices for the remaining parameters in numerical studies are identical to the ones summarized in \cite{Beneke:2020fot}.

%%%%%%%%%%%%%%%%%%%%%%%%%%%%%%%%%%%%%%%%%%%%%%%%%%%%%%%%%%%%%%%
\begin{figure}[tp]
\begin{center}
\includegraphics[width=0.95 \columnwidth]{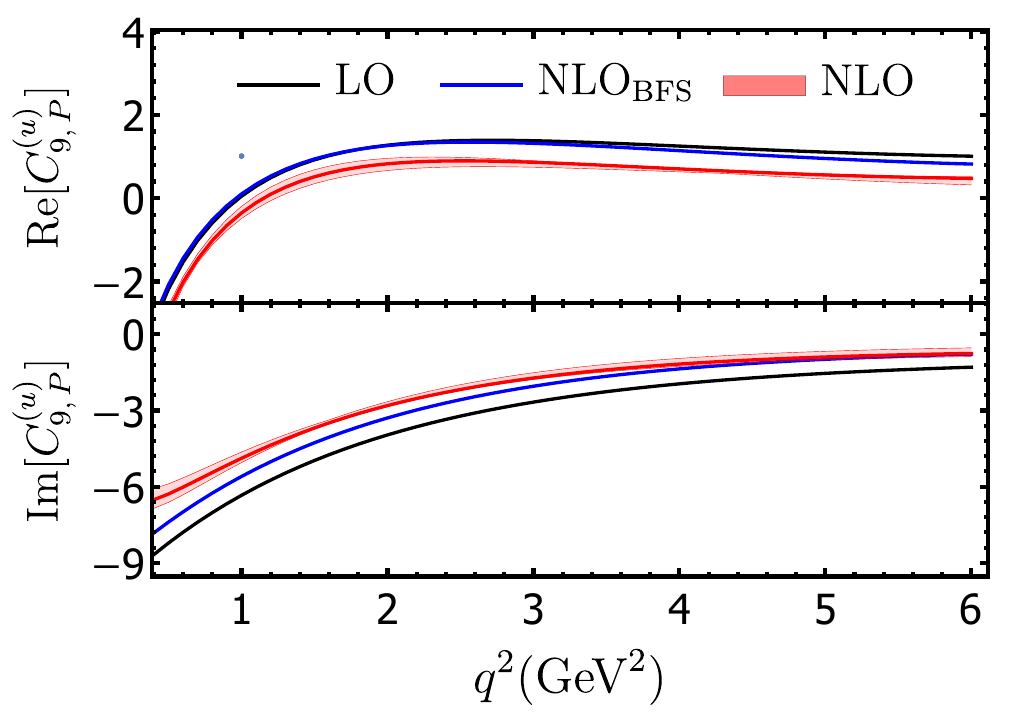} \\
%\vspace*{0.2 cm}
%\includegraphics[width=1.0 \columnwidth]{C9t-pion.pdf}
\vspace*{0.1cm}
\caption{Theory prediction for the $q^2$ dependence of the  effective Wilson coefficient ${\cal C}_{9, P}^{(u)}$
dictating the exclusive $B^{-} \to \pi^{-} \ell^{+} \ell^{-}$ decay amplitude at LO (black curves),
at ${\rm NLO}_{\rm BFS}$ (blue curves) and at NLO (pink bands),
where the notation ``${\rm NLO}_{\rm BFS}$" represents the available NLO computation accomplished in \cite{Beneke:2001at,Beneke:2004dp},
and the uncertainties  are obtained from varying the hard and hard-collinear matching scales. }
\label{fig: Effective Wilson Coefficients C9}
\end{center}
\end{figure}
%%%%%%%%%%%%%%%%%%%%%%%%%%%%%%%%%%%%%%%%%%%%%%%%%%%%%%%%%%%%%%%

Inspecting the numerical features of the effective Wilson coefficient ${\cal C}_{9, P}^{(u)}$
for $B^{-} \to \pi^{-} \ell^{+} \ell^{-}$  displayed in Figure \ref{fig: Effective Wilson Coefficients C9}
indicates that the newly computed NLO weak annihilation contribution can bring about ${\cal O} (35 \, \%)$
(${\cal O} (15 \, \%)$) reduction of the tree-level prediction for the real (imaginary) part of ${\cal C}_{9, P}^{(u)}$
in the kinematic range $1.5 \, {\rm GeV}^2 \leq q^2 \leq 4.0  \, {\rm GeV}^2$.
By contrast, the previously determined NLO QCD correction to the invariant amplitude $\mathcal{T}_P^{(u)}$
(shown in Figure 2 of \cite{Beneke:2001at}) generates, on the one hand,  an  impact on
${\rm Re} [{\cal C}_{9, P}^{(u)}]$  no more  than $10 \, \%$ numerically in the same kinematic region and,
on the other hand, approximately $(15-30) \%$ reduction
of the LO prediction of  ${\rm Im} [{\cal C}_{9, P}^{(u)}]$.
In consequence, the higher-order weak annihilation effect under discussion provides an intriguing source
of the strong phases of the exclusive $b \to d \ell^{+} \ell^{-}$ decay amplitudes.
In addition, the NLO weak annihilation correction to the invariant amplitude $\mathcal{T}_P^{(t)}$
appears to be merely ${\cal O}(3 \, \%)$ of
the previously computed NLO contribution \cite{Beneke:2001at,Beneke:2004dp} in magnitude
at $q^2 \in [1.5, \, 4.0]\, {\rm GeV}^2$, due to the absence of an important contribution from
the current-current operators ${\cal Q}_{1, 2}^{c}$ with the large Wilson coefficients.

%%%%%%%%%%%%%%%%%%%%%%%%%%%%%%%%%%%%%%%%%%%%%%%%%%%%%%%%%%%%%%%
\begin{figure}[tp]
\begin{center}
\includegraphics[width=0.95 \columnwidth]{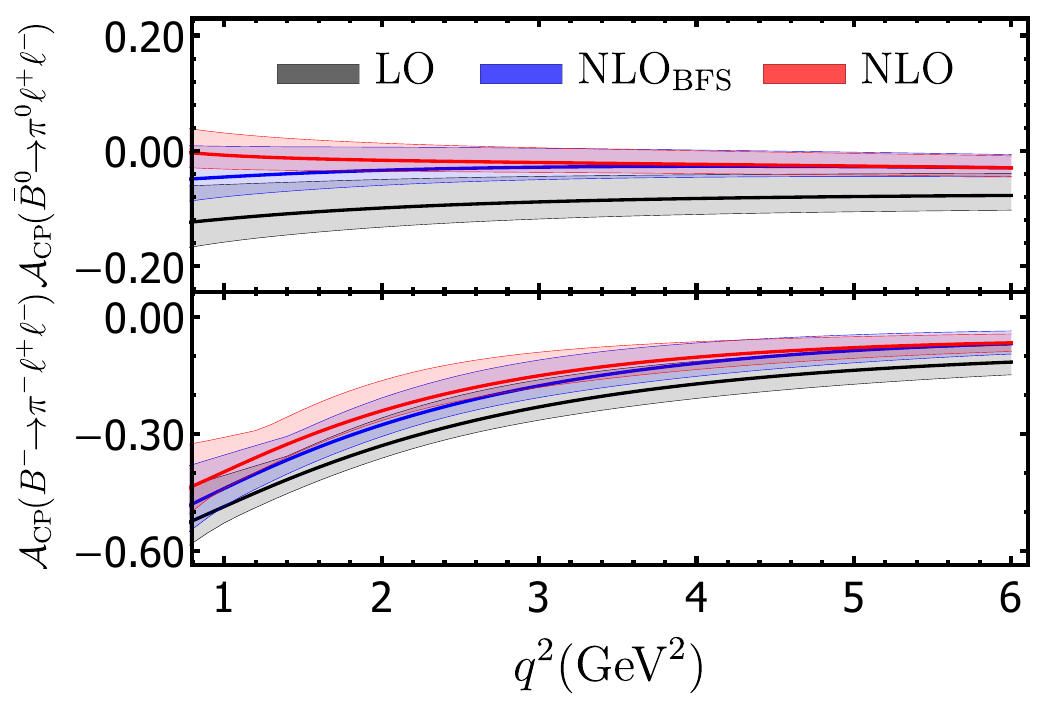} \\
\vspace*{0.5 cm}
\includegraphics[width=0.95 \columnwidth]{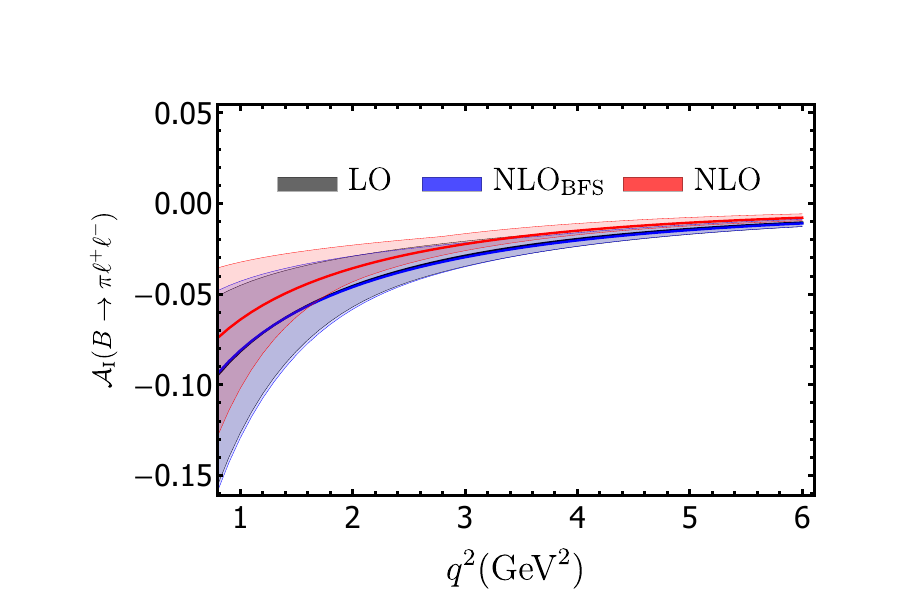}
\vspace*{0.1cm}
\caption{Theory predictions for the $q^2$ dependence of the  direct CP asymmetries [top] and the isospin asymmetry [down]
for the exclusive $B  \to \pi  \ell^{+} \ell^{-}$ decays at LO (grey bands),
at ${\rm NLO}_{\rm BFS}$ (blue bands) and at NLO (pink bands),
where the uncertainties  are obtained by adding the separate errors due to the variations of all input parameters in quadrature. }
\label{fig: Asymmetries for B to pi ll}
\end{center}
\end{figure}
%%%%%%%%%%%%%%%%%%%%%%%%%%%%%%%%%%%%%%%%%%%%%%%%%%%%%%%%%%%%%%%

We proceed  to explore  phenomenological impacts of the determined NLO weak annihilation correction
on the CP-averaged branching fraction (${\cal BR}$),  the direct CP asymmetry ($\mathcal{A}_{\rm CP}$)
and the CP-averaged  isospin asymmetry ($\mathcal{A}_{\rm I}$) for $B \to P \ell^{+} \ell^{-}$,
the latter two of which are explicitly defined by
\begin{eqnarray}
\mathcal{A}_{\rm CP} &=& \frac{\Gamma(\bar B \to \bar P \ell^{+} \ell^{-}) - \Gamma(B \to P \ell^{+} \ell^{-})}
{\Gamma(\bar B \to \bar P \ell^{+} \ell^{-}) + \Gamma(B \to P \ell^{+} \ell^{-})}\,,
\nonumber \\
\mathcal{A}_{\rm I} &=&  \frac{\kappa \, \Gamma(\bar B^{0} \to \bar P^{0} \ell^{+} \ell^{-}) - \Gamma(B^{-} \to P^{-} \ell^{+} \ell^{-})}
{\kappa \, \Gamma(\bar B^{0} \to \bar P^{0} \ell^{+} \ell^{-}) + \Gamma(B^{-} \to P^{-} \ell^{+} \ell^{-})} \,,
\hspace{1.0 cm}
\end{eqnarray}
with the factor $\kappa=2, 1$ for $P=\pi^0, K^{0}$.
Unsurprisingly, including the NLO weak annihilation correction in the improved theory predictions
can   lead to the negligible impacts on ${\cal BR} (B \to K \ell^{+} \ell^{-})$ and ${\cal BR} (B \to \pi \ell^{+} \ell^{-})$.
In virtue of  the the smallness of the direct CP asymmetries for the exclusive $b \to s \ell^{+} \ell^{-}$ decay processes in the SM,
we collect here only our predictions for the binned asymmetries
$\mathcal{A}_{\rm CP}[q_{\rm min}^2, q_{\rm max}^2]=\{1.02^{+0.05}_{-0.22}, \, 0.64^{+0.06}_{-0.24}, \,
0.42^{+0.10}_{-0.17}, \,  0.31^{+0.10}_{-0.13}, 0.26^{+0.09}_{-0.10} \} \, \%$
for  $B^{-} \to K^{-} \ell^{+} \ell^{-}$
and $\mathcal{A}_{\rm CP}[q_{\rm min}^2, q_{\rm max}^2]=\{0.01^{+0.08}_{-0.12}, \, 0.04^{+0.07}_{-0.11},
\,  0.06^{+0.07}_{-0.10}, \, 0.08^{+0.07}_{-0.10},  0.09^{+0.07}_{-0.09} \} \, \%$ for $\bar B^{0} \to \bar K^{0} \ell^{+} \ell^{-}$,
with the choices of the five sample $q^2$ bins (following \cite{LHCb:2014mit}) $[q_{\rm min}^2, q_{\rm max}^2] =
\left \{ [1.1, 2.0], \,  [2.0, 3.0], \,  [3.0, 4.0], \,  [4.0, 5.0], \,  [5.0, 6.0] \right \} \, {\rm GeV^2}$.
Moreover, the yielding results for the isospin asymmetry of $B \to K \ell^{+} \ell^{-}$ in the kinematic ranges
$q^2 \in [2.0,  4.0] \, {\rm GeV^2}$ and $q^2 \in [4.0,  6.0] \, {\rm GeV^2}$
are given by $(-0.74)^{+0.21}_{-0.12} \, \%$ and $(-0.44)^{+0.08}_{-0.06} \, \%$, respectively,
whose magnitudes and signs are both compatible with the updated LHCb measurements \cite{LHCb:2014cxe}.
We now present the comparative predictions for the  direct CP asymmetries and the isopsin asymmetry
of $B \to \pi \ell^{+} \ell^{-}$ in Figure \ref{fig: Asymmetries for B to pi ll}.
% where the predicted binned asymmetries  are also shown
%for comparison with the numerical results determined in \cite{Hambrock:2015wka}.
It is then evident that the inclusion of  the NLO weak annihilation contribution can result in
the noticeable shifts of the theory predictions for these asymmetries in the ${\rm NLO}_{\rm BFS}$ approximation:
numerically ${\cal O}(15 \, \%)$ and ${\cal O}(25 \, \%)$ corrections to
$\mathcal{A}_{\rm CP}(B^{-} \to \pi^{-} \ell^{+} \ell^{-})$
and $\mathcal{A}_{\rm I}(B  \to \pi  \ell^{+} \ell^{-})$.
It is  perhaps  worth mentioning that the predicted  CP asymmetry for $\bar B^{0} \to \pi^{0} \ell^{+} \ell^{-}$
turns out to be significantly smaller than the counterpart result for the charged $B$-meson decay channel,
on account of the  hierarchy structure of the  LO annihilation amplitudes
$ [ T_{P, -}^{(0, u)}(\bar B^{0} \to \pi^{0} \ell^{+} \ell^{-}) ]  :
 [ T_{P, -}^{(0, u)}(B^{-} \to \pi^{-} \ell^{+} \ell^{-})]
= \left [ - Q_d \, (C_F \, C_1 + C_2) \right  ] :  \left [ Q_u \, N_c \, C_2 \right ] \approx  1 : 10$.
Importantly,  the obtained prediction of the isopsin breaking effect in  $B  \to \pi  \ell^{+} \ell^{-}$ is approximately
$3-6$ times larger than that in $B  \to K  \ell^{+} \ell^{-}$ for  $q^2 \in [1.5, \, 4.0]\, {\rm GeV}^2$,
thus enabling this asymmetry observable to become the sensitive probe of the Wilson coefficients of
the current-current operators in the weak effective Hamiltonian.

%%%%%%%%%%%%%%%%%%%%%%%%%%%%%%%%%%%%%%%%%%%%%%%%%%%%%%%%%%%%%%%%%%%%%%%%%%%%
\section{Conclusions}
%%%%%%%%%%%%%%%%%%%%%%%%%%%%%%%%%%%%%%%%%%%%%%%%%%%%%%%%%%%%%%%%%%%%%%%%%%%%

In conclusion, we have presented the first computation of the NLO weak annihilation contribution
to the   rare $B \to \left \{K, \pi \right \} \ell^{+} \ell^{-}$ decays
at leading power in the heavy quark expansion by employing the soft-collinear factorization framework,
thus providing the last missing piece in evaluating the QCD correction to
the isospin asymmetry $\mathcal{A}_{\rm I}(B  \to \pi  \ell^{+} \ell^{-})$.
In contrast with the tree-level annihilation amplitude, we have demonstrated that
both the twist-two and twist-three HQET LCDAs of the bottom-meson  are in demand
to capture the soft QCD dynamics in the established factorization formula
of the weak annihilation effect at ${\cal O}(\alpha_s)$.
Taking into account such higher-order corrections in the matrix elements of  the four-quark operators
can result in the notable impacts on our theory  predictions for the direct CP asymmetry
$\mathcal{A}_{\rm CP}(B^{-} \to \pi^{-} \ell^{+} \ell^{-})$
as well as the CP-averaged isospin asymmetry $\mathcal{A}_{\rm I}(B  \to \pi  \ell^{+} \ell^{-})$.
In addition, the achieved  NLO factorization formula of the annihilation amplitude
can be straightforwardly extended to the electroweak penguin decays $B \to \left \{K^{\ast}, \, \rho, \, \omega \right \} \ell^{+} \ell^{-}$
with the longitudinally polarized vector mesons (by implementing  the appropriate replacements),
which are generally accepted as the benchmark channels for exploring  the quark flavour-changing dynamics.

%
%%%%%%%%%%%%%%%%%%%%%%%%%%%%%%%%%%%%%%%%%%%%%%%%%%%%%%%%%%%%%%%%%%%%%%%%%%%%
\begin{acknowledgments}
\section*{Acknowledgements}

Y.K.H. and Y.M.W. acknowledge support from the National Natural Science Foundation of China  with
Grant No.  12075125.
%and the Natural Science Foundation of Tianjin with Grant No. 19JCJQJC61100.
The research of Y.L.S. is supported by the  National Natural Science Foundation of China  with
Grant No. 12175218.
C.W. is supported in part by the National Natural Science Foundation of China
with Grant No. 12105112 and  the Natural Science Foundation of
Jiangsu Education Committee with Grant No. 21KJB140027.

\end{acknowledgments}
%%%%%%%%%%%%%%%%%%%%%%%%%%%%%%%%%%%%%%%%%%%%%%%%%%%%%%%%%%%%%%%%%%%%%%%%%%%%

%%%%%%%%%%%%%%%%%%%%%%%%%%%%%%%%%%%%%%%%%%%%%%%%%%%%%%%%%%%%%%%%%%%%%%%%%%%%
\bibliographystyle{apsrev4-1}

\bibliography{References}

\begin{thebibliography}{75}%
\makeatletter
\providecommand \@ifxundefined [1]{%
 \@ifx{#1\undefined}
}%
\providecommand \@ifnum [1]{%
 \ifnum #1\expandafter \@firstoftwo
 \else \expandafter \@secondoftwo
 \fi
}%
\providecommand \@ifx [1]{%
 \ifx #1\expandafter \@firstoftwo
 \else \expandafter \@secondoftwo
 \fi
}%
\providecommand \natexlab [1]{#1}%
\providecommand \enquote  [1]{``#1''}%
\providecommand \bibnamefont  [1]{#1}%
\providecommand \bibfnamefont [1]{#1}%
\providecommand \citenamefont [1]{#1}%
\providecommand \href@noop [0]{\@secondoftwo}%
\providecommand \href [0]{\begingroup \@sanitize@url \@href}%
\providecommand \@href[1]{\@@startlink{#1}\@@href}%
\providecommand \@@href[1]{\endgroup#1\@@endlink}%
\providecommand \@sanitize@url [0]{\catcode `\\12\catcode `\$12\catcode
  `\&12\catcode `\#12\catcode `\^12\catcode `\_12\catcode `\%12\relax}%
\providecommand \@@startlink[1]{}%
\providecommand \@@endlink[0]{}%
\providecommand \url  [0]{\begingroup\@sanitize@url \@url }%
\providecommand \@url [1]{\endgroup\@href {#1}{\urlprefix }}%
\providecommand \urlprefix  [0]{URL }%
\providecommand \Eprint [0]{\href }%
\providecommand \doibase [0]{http://dx.doi.org/}%
\providecommand \selectlanguage [0]{\@gobble}%
\providecommand \bibinfo  [0]{\@secondoftwo}%
\providecommand \bibfield  [0]{\@secondoftwo}%
\providecommand \translation [1]{[#1]}%
\providecommand \BibitemOpen [0]{}%
\providecommand \bibitemStop [0]{}%
\providecommand \bibitemNoStop [0]{.\EOS\space}%
\providecommand \EOS [0]{\spacefactor3000\relax}%
\providecommand \BibitemShut  [1]{\csname bibitem#1\endcsname}%
\let\auto@bib@innerbib\@empty
%</preamble>
\bibitem [{\citenamefont {Capdevila}\ \emph {et~al.}(2023)\citenamefont
  {Capdevila}, \citenamefont {Crivellin},\ and\ \citenamefont
  {Matias}}]{Capdevila:2023yhq}%
  \BibitemOpen
  \bibfield  {author} {\bibinfo {author} {\bibfnamefont {B.}~\bibnamefont
  {Capdevila}}, \bibinfo {author} {\bibfnamefont {A.}~\bibnamefont
  {Crivellin}}, \ and\ \bibinfo {author} {\bibfnamefont {J.}~\bibnamefont
  {Matias}},\ }\href {\doibase 10.1140/epjs/s11734-023-01012-2} {\bibfield
  {journal} {\bibinfo  {journal} {Eur. Phys. J. ST}\ }\textbf {\bibinfo
  {volume} {1}},\ \bibinfo {pages} {20} (\bibinfo {year} {2023})},\ \Eprint
  {http://arxiv.org/abs/2309.01311} {arXiv:2309.01311 [hep-ph]} \BibitemShut
  {NoStop}%
\bibitem [{\citenamefont {Aaij}\ \emph {et~al.}(2012)\citenamefont {Aaij} \emph
  {et~al.}}]{LHCb:2012bin}%
  \BibitemOpen
  \bibfield  {author} {\bibinfo {author} {\bibfnamefont {R.}~\bibnamefont
  {Aaij}} \emph {et~al.} (\bibinfo {collaboration} {LHCb}),\ }\href {\doibase
  10.1007/JHEP07(2012)133} {\bibfield  {journal} {\bibinfo  {journal} {JHEP}\
  }\textbf {\bibinfo {volume} {07}},\ \bibinfo {pages} {133} (\bibinfo {year}
  {2012})},\ \Eprint {http://arxiv.org/abs/1205.3422} {arXiv:1205.3422
  [hep-ex]} \BibitemShut {NoStop}%
\bibitem [{\citenamefont {Aaij}\ \emph
  {et~al.}(2014{\natexlab{a}})\citenamefont {Aaij} \emph
  {et~al.}}]{LHCb:2014cxe}%
  \BibitemOpen
  \bibfield  {author} {\bibinfo {author} {\bibfnamefont {R.}~\bibnamefont
  {Aaij}} \emph {et~al.} (\bibinfo {collaboration} {LHCb}),\ }\href {\doibase
  10.1007/JHEP06(2014)133} {\bibfield  {journal} {\bibinfo  {journal} {JHEP}\
  }\textbf {\bibinfo {volume} {06}},\ \bibinfo {pages} {133} (\bibinfo {year}
  {2014}{\natexlab{a}})},\ \Eprint {http://arxiv.org/abs/1403.8044}
  {arXiv:1403.8044 [hep-ex]} \BibitemShut {NoStop}%
\bibitem [{\citenamefont {Feldmann}\ and\ \citenamefont
  {Matias}(2003)}]{Feldmann:2002iw}%
  \BibitemOpen
  \bibfield  {author} {\bibinfo {author} {\bibfnamefont {T.}~\bibnamefont
  {Feldmann}}\ and\ \bibinfo {author} {\bibfnamefont {J.}~\bibnamefont
  {Matias}},\ }\href {\doibase 10.1088/1126-6708/2003/01/074} {\bibfield
  {journal} {\bibinfo  {journal} {JHEP}\ }\textbf {\bibinfo {volume} {01}},\
  \bibinfo {pages} {074} (\bibinfo {year} {2003})},\ \Eprint
  {http://arxiv.org/abs/hep-ph/0212158} {arXiv:hep-ph/0212158} \BibitemShut
  {NoStop}%
\bibitem [{\citenamefont {Khodjamirian}\ \emph {et~al.}(2013)\citenamefont
  {Khodjamirian}, \citenamefont {Mannel},\ and\ \citenamefont
  {Wang}}]{Khodjamirian:2012rm}%
  \BibitemOpen
  \bibfield  {author} {\bibinfo {author} {\bibfnamefont {A.}~\bibnamefont
  {Khodjamirian}}, \bibinfo {author} {\bibfnamefont {T.}~\bibnamefont
  {Mannel}}, \ and\ \bibinfo {author} {\bibfnamefont {Y.~M.}\ \bibnamefont
  {Wang}},\ }\href {\doibase 10.1007/JHEP02(2013)010} {\bibfield  {journal}
  {\bibinfo  {journal} {JHEP}\ }\textbf {\bibinfo {volume} {02}},\ \bibinfo
  {pages} {010} (\bibinfo {year} {2013})},\ \Eprint
  {http://arxiv.org/abs/1211.0234} {arXiv:1211.0234 [hep-ph]} \BibitemShut
  {NoStop}%
\bibitem [{\citenamefont {Lyon}\ and\ \citenamefont
  {Zwicky}(2013)}]{Lyon:2013gba}%
  \BibitemOpen
  \bibfield  {author} {\bibinfo {author} {\bibfnamefont {J.}~\bibnamefont
  {Lyon}}\ and\ \bibinfo {author} {\bibfnamefont {R.}~\bibnamefont {Zwicky}},\
  }\href {\doibase 10.1103/PhysRevD.88.094004} {\bibfield  {journal} {\bibinfo
  {journal} {Phys. Rev. D}\ }\textbf {\bibinfo {volume} {88}},\ \bibinfo
  {pages} {094004} (\bibinfo {year} {2013})},\ \Eprint
  {http://arxiv.org/abs/1305.4797} {arXiv:1305.4797 [hep-ph]} \BibitemShut
  {NoStop}%
\bibitem [{\citenamefont {Hou}\ \emph {et~al.}(2014)\citenamefont {Hou},
  \citenamefont {Kohda},\ and\ \citenamefont {Xu}}]{Hou:2014dza}%
  \BibitemOpen
  \bibfield  {author} {\bibinfo {author} {\bibfnamefont {W.-S.}\ \bibnamefont
  {Hou}}, \bibinfo {author} {\bibfnamefont {M.}~\bibnamefont {Kohda}}, \ and\
  \bibinfo {author} {\bibfnamefont {F.}~\bibnamefont {Xu}},\ }\href {\doibase
  10.1103/PhysRevD.90.013002} {\bibfield  {journal} {\bibinfo  {journal} {Phys.
  Rev. D}\ }\textbf {\bibinfo {volume} {90}},\ \bibinfo {pages} {013002}
  (\bibinfo {year} {2014})},\ \Eprint {http://arxiv.org/abs/1403.7410}
  {arXiv:1403.7410 [hep-ph]} \BibitemShut {NoStop}%
\bibitem [{\citenamefont {Hambrock}\ \emph {et~al.}(2015)\citenamefont
  {Hambrock}, \citenamefont {Khodjamirian},\ and\ \citenamefont
  {Rusov}}]{Hambrock:2015wka}%
  \BibitemOpen
  \bibfield  {author} {\bibinfo {author} {\bibfnamefont {C.}~\bibnamefont
  {Hambrock}}, \bibinfo {author} {\bibfnamefont {A.}~\bibnamefont
  {Khodjamirian}}, \ and\ \bibinfo {author} {\bibfnamefont {A.}~\bibnamefont
  {Rusov}},\ }\href {\doibase 10.1103/PhysRevD.92.074020} {\bibfield  {journal}
  {\bibinfo  {journal} {Phys. Rev. D}\ }\textbf {\bibinfo {volume} {92}},\
  \bibinfo {pages} {074020} (\bibinfo {year} {2015})},\ \Eprint
  {http://arxiv.org/abs/1506.07760} {arXiv:1506.07760 [hep-ph]} \BibitemShut
  {NoStop}%
\bibitem [{\citenamefont {Kagan}\ and\ \citenamefont
  {Neubert}(2002)}]{Kagan:2001zk}%
  \BibitemOpen
  \bibfield  {author} {\bibinfo {author} {\bibfnamefont {A.~L.}\ \bibnamefont
  {Kagan}}\ and\ \bibinfo {author} {\bibfnamefont {M.}~\bibnamefont
  {Neubert}},\ }\href {\doibase 10.1016/S0370-2693(02)02100-7} {\bibfield
  {journal} {\bibinfo  {journal} {Phys. Lett. B}\ }\textbf {\bibinfo {volume}
  {539}},\ \bibinfo {pages} {227} (\bibinfo {year} {2002})},\ \Eprint
  {http://arxiv.org/abs/hep-ph/0110078} {arXiv:hep-ph/0110078} \BibitemShut
  {NoStop}%
\bibitem [{\citenamefont {Ball}\ \emph {et~al.}(2007)\citenamefont {Ball},
  \citenamefont {Jones},\ and\ \citenamefont {Zwicky}}]{Ball:2006eu}%
  \BibitemOpen
  \bibfield  {author} {\bibinfo {author} {\bibfnamefont {P.}~\bibnamefont
  {Ball}}, \bibinfo {author} {\bibfnamefont {G.~W.}\ \bibnamefont {Jones}}, \
  and\ \bibinfo {author} {\bibfnamefont {R.}~\bibnamefont {Zwicky}},\ }\href
  {\doibase 10.1103/PhysRevD.75.054004} {\bibfield  {journal} {\bibinfo
  {journal} {Phys. Rev. D}\ }\textbf {\bibinfo {volume} {75}},\ \bibinfo
  {pages} {054004} (\bibinfo {year} {2007})},\ \Eprint
  {http://arxiv.org/abs/hep-ph/0612081} {arXiv:hep-ph/0612081} \BibitemShut
  {NoStop}%
\bibitem [{\citenamefont {Khodjamirian}\ \emph {et~al.}(2010)\citenamefont
  {Khodjamirian}, \citenamefont {Mannel}, \citenamefont {Pivovarov},\ and\
  \citenamefont {Wang}}]{Khodjamirian:2010vf}%
  \BibitemOpen
  \bibfield  {author} {\bibinfo {author} {\bibfnamefont {A.}~\bibnamefont
  {Khodjamirian}}, \bibinfo {author} {\bibfnamefont {T.}~\bibnamefont
  {Mannel}}, \bibinfo {author} {\bibfnamefont {A.~A.}\ \bibnamefont
  {Pivovarov}}, \ and\ \bibinfo {author} {\bibfnamefont {Y.~M.}\ \bibnamefont
  {Wang}},\ }\href {\doibase 10.1007/JHEP09(2010)089} {\bibfield  {journal}
  {\bibinfo  {journal} {JHEP}\ }\textbf {\bibinfo {volume} {09}},\ \bibinfo
  {pages} {089} (\bibinfo {year} {2010})},\ \Eprint
  {http://arxiv.org/abs/1006.4945} {arXiv:1006.4945 [hep-ph]} \BibitemShut
  {NoStop}%
\bibitem [{\citenamefont {Gubernari}\ \emph {et~al.}(2021)\citenamefont
  {Gubernari}, \citenamefont {van Dyk},\ and\ \citenamefont
  {Virto}}]{Gubernari:2020eft}%
  \BibitemOpen
  \bibfield  {author} {\bibinfo {author} {\bibfnamefont {N.}~\bibnamefont
  {Gubernari}}, \bibinfo {author} {\bibfnamefont {D.}~\bibnamefont {van Dyk}},
  \ and\ \bibinfo {author} {\bibfnamefont {J.}~\bibnamefont {Virto}},\ }\href
  {\doibase 10.1007/JHEP02(2021)088} {\bibfield  {journal} {\bibinfo  {journal}
  {JHEP}\ }\textbf {\bibinfo {volume} {02}},\ \bibinfo {pages} {088} (\bibinfo
  {year} {2021})},\ \Eprint {http://arxiv.org/abs/2011.09813} {arXiv:2011.09813
  [hep-ph]} \BibitemShut {NoStop}%
\bibitem [{\citenamefont {Bobeth}\ \emph {et~al.}(2018)\citenamefont {Bobeth},
  \citenamefont {Chrzaszcz}, \citenamefont {van Dyk},\ and\ \citenamefont
  {Virto}}]{Bobeth:2017vxj}%
  \BibitemOpen
  \bibfield  {author} {\bibinfo {author} {\bibfnamefont {C.}~\bibnamefont
  {Bobeth}}, \bibinfo {author} {\bibfnamefont {M.}~\bibnamefont {Chrzaszcz}},
  \bibinfo {author} {\bibfnamefont {D.}~\bibnamefont {van Dyk}}, \ and\
  \bibinfo {author} {\bibfnamefont {J.}~\bibnamefont {Virto}},\ }\href
  {\doibase 10.1140/epjc/s10052-018-5918-6} {\bibfield  {journal} {\bibinfo
  {journal} {Eur. Phys. J. C}\ }\textbf {\bibinfo {volume} {78}},\ \bibinfo
  {pages} {451} (\bibinfo {year} {2018})},\ \Eprint
  {http://arxiv.org/abs/1707.07305} {arXiv:1707.07305 [hep-ph]} \BibitemShut
  {NoStop}%
\bibitem [{\citenamefont {Bharucha}\ \emph {et~al.}(2016)\citenamefont
  {Bharucha}, \citenamefont {Straub},\ and\ \citenamefont
  {Zwicky}}]{Bharucha:2015bzk}%
  \BibitemOpen
  \bibfield  {author} {\bibinfo {author} {\bibfnamefont {A.}~\bibnamefont
  {Bharucha}}, \bibinfo {author} {\bibfnamefont {D.~M.}\ \bibnamefont
  {Straub}}, \ and\ \bibinfo {author} {\bibfnamefont {R.}~\bibnamefont
  {Zwicky}},\ }\href {\doibase 10.1007/JHEP08(2016)098} {\bibfield  {journal}
  {\bibinfo  {journal} {JHEP}\ }\textbf {\bibinfo {volume} {08}},\ \bibinfo
  {pages} {098} (\bibinfo {year} {2016})},\ \Eprint
  {http://arxiv.org/abs/1503.05534} {arXiv:1503.05534 [hep-ph]} \BibitemShut
  {NoStop}%
\bibitem [{\citenamefont {Gao}\ \emph {et~al.}(2020)\citenamefont {Gao},
  \citenamefont {L\"u}, \citenamefont {Shen}, \citenamefont {Wang},\ and\
  \citenamefont {Wei}}]{Gao:2019lta}%
  \BibitemOpen
  \bibfield  {author} {\bibinfo {author} {\bibfnamefont {J.}~\bibnamefont
  {Gao}}, \bibinfo {author} {\bibfnamefont {C.-D.}\ \bibnamefont {L\"u}},
  \bibinfo {author} {\bibfnamefont {Y.-L.}\ \bibnamefont {Shen}}, \bibinfo
  {author} {\bibfnamefont {Y.-M.}\ \bibnamefont {Wang}}, \ and\ \bibinfo
  {author} {\bibfnamefont {Y.-B.}\ \bibnamefont {Wei}},\ }\href {\doibase
  10.1103/PhysRevD.101.074035} {\bibfield  {journal} {\bibinfo  {journal}
  {Phys. Rev. D}\ }\textbf {\bibinfo {volume} {101}},\ \bibinfo {pages}
  {074035} (\bibinfo {year} {2020})},\ \Eprint
  {http://arxiv.org/abs/1907.11092} {arXiv:1907.11092 [hep-ph]} \BibitemShut
  {NoStop}%
\bibitem [{\citenamefont {Descotes-Genon}\ \emph {et~al.}(2023)\citenamefont
  {Descotes-Genon}, \citenamefont {Khodjamirian}, \citenamefont {Virto},\ and\
  \citenamefont {Vos}}]{Descotes-Genon:2023ukb}%
  \BibitemOpen
  \bibfield  {author} {\bibinfo {author} {\bibfnamefont {S.}~\bibnamefont
  {Descotes-Genon}}, \bibinfo {author} {\bibfnamefont {A.}~\bibnamefont
  {Khodjamirian}}, \bibinfo {author} {\bibfnamefont {J.}~\bibnamefont {Virto}},
  \ and\ \bibinfo {author} {\bibfnamefont {K.~K.}\ \bibnamefont {Vos}},\ }\href
  {\doibase 10.1007/JHEP06(2023)034} {\bibfield  {journal} {\bibinfo  {journal}
  {JHEP}\ }\textbf {\bibinfo {volume} {06}},\ \bibinfo {pages} {034} (\bibinfo
  {year} {2023})},\ \Eprint {http://arxiv.org/abs/2304.02973} {arXiv:2304.02973
  [hep-ph]} \BibitemShut {NoStop}%
\bibitem [{\citenamefont {L\"u}\ \emph {et~al.}(2019)\citenamefont {L\"u},
  \citenamefont {Shen}, \citenamefont {Wang},\ and\ \citenamefont
  {Wei}}]{Lu:2018cfc}%
  \BibitemOpen
  \bibfield  {author} {\bibinfo {author} {\bibfnamefont {C.-D.}\ \bibnamefont
  {L\"u}}, \bibinfo {author} {\bibfnamefont {Y.-L.}\ \bibnamefont {Shen}},
  \bibinfo {author} {\bibfnamefont {Y.-M.}\ \bibnamefont {Wang}}, \ and\
  \bibinfo {author} {\bibfnamefont {Y.-B.}\ \bibnamefont {Wei}},\ }\href
  {\doibase 10.1007/JHEP01(2019)024} {\bibfield  {journal} {\bibinfo  {journal}
  {JHEP}\ }\textbf {\bibinfo {volume} {01}},\ \bibinfo {pages} {024} (\bibinfo
  {year} {2019})},\ \Eprint {http://arxiv.org/abs/1810.00819} {arXiv:1810.00819
  [hep-ph]} \BibitemShut {NoStop}%
\bibitem [{\citenamefont {Cui}\ \emph {et~al.}(2023)\citenamefont {Cui},
  \citenamefont {Huang}, \citenamefont {Shen}, \citenamefont {Wang},\ and\
  \citenamefont {Wang}}]{Cui:2022zwm}%
  \BibitemOpen
  \bibfield  {author} {\bibinfo {author} {\bibfnamefont {B.-Y.}\ \bibnamefont
  {Cui}}, \bibinfo {author} {\bibfnamefont {Y.-K.}\ \bibnamefont {Huang}},
  \bibinfo {author} {\bibfnamefont {Y.-L.}\ \bibnamefont {Shen}}, \bibinfo
  {author} {\bibfnamefont {C.}~\bibnamefont {Wang}}, \ and\ \bibinfo {author}
  {\bibfnamefont {Y.-M.}\ \bibnamefont {Wang}},\ }\href {\doibase
  10.1007/JHEP03(2023)140} {\bibfield  {journal} {\bibinfo  {journal} {JHEP}\
  }\textbf {\bibinfo {volume} {03}},\ \bibinfo {pages} {140} (\bibinfo {year}
  {2023})},\ \Eprint {http://arxiv.org/abs/2212.11624} {arXiv:2212.11624
  [hep-ph]} \BibitemShut {NoStop}%
\bibitem [{\citenamefont {Huang}\ \emph {et~al.}(2023)\citenamefont {Huang},
  \citenamefont {Ji}, \citenamefont {Shen}, \citenamefont {Wang}, \citenamefont
  {Wang},\ and\ \citenamefont {Zhao}}]{Huang:2023jdu}%
  \BibitemOpen
  \bibfield  {author} {\bibinfo {author} {\bibfnamefont {Y.-K.}\ \bibnamefont
  {Huang}}, \bibinfo {author} {\bibfnamefont {Y.}~\bibnamefont {Ji}}, \bibinfo
  {author} {\bibfnamefont {Y.-L.}\ \bibnamefont {Shen}}, \bibinfo {author}
  {\bibfnamefont {C.}~\bibnamefont {Wang}}, \bibinfo {author} {\bibfnamefont
  {Y.-M.}\ \bibnamefont {Wang}}, \ and\ \bibinfo {author} {\bibfnamefont
  {X.-C.}\ \bibnamefont {Zhao}},\ }\href@noop {} {\  (\bibinfo {year}
  {2023})},\ \Eprint {http://arxiv.org/abs/2312.15439} {arXiv:2312.15439
  [hep-ph]} \BibitemShut {NoStop}%
\bibitem [{\citenamefont {Khodjamirian}\ \emph {et~al.}(2023)\citenamefont
  {Khodjamirian}, \citenamefont {Meli\'c},\ and\ \citenamefont
  {Wang}}]{Khodjamirian:2023wol}%
  \BibitemOpen
  \bibfield  {author} {\bibinfo {author} {\bibfnamefont {A.}~\bibnamefont
  {Khodjamirian}}, \bibinfo {author} {\bibfnamefont {B.}~\bibnamefont
  {Meli\'c}}, \ and\ \bibinfo {author} {\bibfnamefont {Y.-M.}\ \bibnamefont
  {Wang}},\ }\href {\doibase 10.1140/epjs/s11734-023-01046-6} {\  (\bibinfo
  {year} {2023}),\ 10.1140/epjs/s11734-023-01046-6},\ \Eprint
  {http://arxiv.org/abs/2311.08700} {arXiv:2311.08700 [hep-ph]} \BibitemShut
  {NoStop}%
\bibitem [{\citenamefont {Beneke}\ \emph {et~al.}(2001)\citenamefont {Beneke},
  \citenamefont {Feldmann},\ and\ \citenamefont {Seidel}}]{Beneke:2001at}%
  \BibitemOpen
  \bibfield  {author} {\bibinfo {author} {\bibfnamefont {M.}~\bibnamefont
  {Beneke}}, \bibinfo {author} {\bibfnamefont {T.}~\bibnamefont {Feldmann}}, \
  and\ \bibinfo {author} {\bibfnamefont {D.}~\bibnamefont {Seidel}},\ }\href
  {\doibase 10.1016/S0550-3213(01)00366-2} {\bibfield  {journal} {\bibinfo
  {journal} {Nucl. Phys. B}\ }\textbf {\bibinfo {volume} {612}},\ \bibinfo
  {pages} {25} (\bibinfo {year} {2001})},\ \Eprint
  {http://arxiv.org/abs/hep-ph/0106067} {arXiv:hep-ph/0106067} \BibitemShut
  {NoStop}%
\bibitem [{\citenamefont {Beneke}\ \emph {et~al.}(2005)\citenamefont {Beneke},
  \citenamefont {Feldmann},\ and\ \citenamefont {Seidel}}]{Beneke:2004dp}%
  \BibitemOpen
  \bibfield  {author} {\bibinfo {author} {\bibfnamefont {M.}~\bibnamefont
  {Beneke}}, \bibinfo {author} {\bibfnamefont {T.}~\bibnamefont {Feldmann}}, \
  and\ \bibinfo {author} {\bibfnamefont {D.}~\bibnamefont {Seidel}},\ }\href
  {\doibase 10.1140/epjc/s2005-02181-5} {\bibfield  {journal} {\bibinfo
  {journal} {Eur. Phys. J. C}\ }\textbf {\bibinfo {volume} {41}},\ \bibinfo
  {pages} {173} (\bibinfo {year} {2005})},\ \Eprint
  {http://arxiv.org/abs/hep-ph/0412400} {arXiv:hep-ph/0412400} \BibitemShut
  {NoStop}%
\bibitem [{\citenamefont {Becher}\ \emph {et~al.}(2005)\citenamefont {Becher},
  \citenamefont {Hill},\ and\ \citenamefont {Neubert}}]{Becher:2005fg}%
  \BibitemOpen
  \bibfield  {author} {\bibinfo {author} {\bibfnamefont {T.}~\bibnamefont
  {Becher}}, \bibinfo {author} {\bibfnamefont {R.~J.}\ \bibnamefont {Hill}}, \
  and\ \bibinfo {author} {\bibfnamefont {M.}~\bibnamefont {Neubert}},\ }\href
  {\doibase 10.1103/PhysRevD.72.094017} {\bibfield  {journal} {\bibinfo
  {journal} {Phys. Rev. D}\ }\textbf {\bibinfo {volume} {72}},\ \bibinfo
  {pages} {094017} (\bibinfo {year} {2005})},\ \Eprint
  {http://arxiv.org/abs/hep-ph/0503263} {arXiv:hep-ph/0503263} \BibitemShut
  {NoStop}%
\bibitem [{\citenamefont {Ali}\ \emph {et~al.}(2006)\citenamefont {Ali},
  \citenamefont {Kramer},\ and\ \citenamefont {Zhu}}]{Ali:2006ew}%
  \BibitemOpen
  \bibfield  {author} {\bibinfo {author} {\bibfnamefont {A.}~\bibnamefont
  {Ali}}, \bibinfo {author} {\bibfnamefont {G.}~\bibnamefont {Kramer}}, \ and\
  \bibinfo {author} {\bibfnamefont {G.-h.}\ \bibnamefont {Zhu}},\ }\href
  {\doibase 10.1140/epjc/s2006-02596-4} {\bibfield  {journal} {\bibinfo
  {journal} {Eur. Phys. J. C}\ }\textbf {\bibinfo {volume} {47}},\ \bibinfo
  {pages} {625} (\bibinfo {year} {2006})},\ \Eprint
  {http://arxiv.org/abs/hep-ph/0601034} {arXiv:hep-ph/0601034} \BibitemShut
  {NoStop}%
\bibitem [{\citenamefont {Beneke}\ \emph {et~al.}(1999)\citenamefont {Beneke},
  \citenamefont {Buchalla}, \citenamefont {Neubert},\ and\ \citenamefont
  {Sachrajda}}]{Beneke:1999br}%
  \BibitemOpen
  \bibfield  {author} {\bibinfo {author} {\bibfnamefont {M.}~\bibnamefont
  {Beneke}}, \bibinfo {author} {\bibfnamefont {G.}~\bibnamefont {Buchalla}},
  \bibinfo {author} {\bibfnamefont {M.}~\bibnamefont {Neubert}}, \ and\
  \bibinfo {author} {\bibfnamefont {C.~T.}\ \bibnamefont {Sachrajda}},\ }\href
  {\doibase 10.1103/PhysRevLett.83.1914} {\bibfield  {journal} {\bibinfo
  {journal} {Phys. Rev. Lett.}\ }\textbf {\bibinfo {volume} {83}},\ \bibinfo
  {pages} {1914} (\bibinfo {year} {1999})},\ \Eprint
  {http://arxiv.org/abs/hep-ph/9905312} {arXiv:hep-ph/9905312} \BibitemShut
  {NoStop}%
\bibitem [{\citenamefont {Beneke}\ \emph {et~al.}(2000)\citenamefont {Beneke},
  \citenamefont {Buchalla}, \citenamefont {Neubert},\ and\ \citenamefont
  {Sachrajda}}]{Beneke:2000ry}%
  \BibitemOpen
  \bibfield  {author} {\bibinfo {author} {\bibfnamefont {M.}~\bibnamefont
  {Beneke}}, \bibinfo {author} {\bibfnamefont {G.}~\bibnamefont {Buchalla}},
  \bibinfo {author} {\bibfnamefont {M.}~\bibnamefont {Neubert}}, \ and\
  \bibinfo {author} {\bibfnamefont {C.~T.}\ \bibnamefont {Sachrajda}},\ }\href
  {\doibase 10.1016/S0550-3213(00)00559-9} {\bibfield  {journal} {\bibinfo
  {journal} {Nucl. Phys. B}\ }\textbf {\bibinfo {volume} {591}},\ \bibinfo
  {pages} {313} (\bibinfo {year} {2000})},\ \Eprint
  {http://arxiv.org/abs/hep-ph/0006124} {arXiv:hep-ph/0006124} \BibitemShut
  {NoStop}%
\bibitem [{\citenamefont {Grinstein}\ and\ \citenamefont
  {Pirjol}(2000)}]{Grinstein:2000pc}%
  \BibitemOpen
  \bibfield  {author} {\bibinfo {author} {\bibfnamefont {B.}~\bibnamefont
  {Grinstein}}\ and\ \bibinfo {author} {\bibfnamefont {D.}~\bibnamefont
  {Pirjol}},\ }\href {\doibase 10.1103/PhysRevD.62.093002} {\bibfield
  {journal} {\bibinfo  {journal} {Phys. Rev. D}\ }\textbf {\bibinfo {volume}
  {62}},\ \bibinfo {pages} {093002} (\bibinfo {year} {2000})},\ \Eprint
  {http://arxiv.org/abs/hep-ph/0002216} {arXiv:hep-ph/0002216} \BibitemShut
  {NoStop}%
\bibitem [{\citenamefont {Beneke}\ and\ \citenamefont
  {Feldmann}(2001)}]{Beneke:2000wa}%
  \BibitemOpen
  \bibfield  {author} {\bibinfo {author} {\bibfnamefont {M.}~\bibnamefont
  {Beneke}}\ and\ \bibinfo {author} {\bibfnamefont {T.}~\bibnamefont
  {Feldmann}},\ }\href {\doibase 10.1016/S0550-3213(00)00585-X} {\bibfield
  {journal} {\bibinfo  {journal} {Nucl. Phys. B}\ }\textbf {\bibinfo {volume}
  {592}},\ \bibinfo {pages} {3} (\bibinfo {year} {2001})},\ \Eprint
  {http://arxiv.org/abs/hep-ph/0008255} {arXiv:hep-ph/0008255} \BibitemShut
  {NoStop}%
\bibitem [{\citenamefont {Grozin}\ and\ \citenamefont
  {Neubert}(1997)}]{Grozin:1996pq}%
  \BibitemOpen
  \bibfield  {author} {\bibinfo {author} {\bibfnamefont {A.~G.}\ \bibnamefont
  {Grozin}}\ and\ \bibinfo {author} {\bibfnamefont {M.}~\bibnamefont
  {Neubert}},\ }\href {\doibase 10.1103/PhysRevD.55.272} {\bibfield  {journal}
  {\bibinfo  {journal} {Phys. Rev. D}\ }\textbf {\bibinfo {volume} {55}},\
  \bibinfo {pages} {272} (\bibinfo {year} {1997})},\ \Eprint
  {http://arxiv.org/abs/hep-ph/9607366} {arXiv:hep-ph/9607366} \BibitemShut
  {NoStop}%
\bibitem [{\citenamefont {Lepage}\ and\ \citenamefont
  {Brodsky}(1979)}]{Lepage:1979zb}%
  \BibitemOpen
  \bibfield  {author} {\bibinfo {author} {\bibfnamefont {G.~P.}\ \bibnamefont
  {Lepage}}\ and\ \bibinfo {author} {\bibfnamefont {S.~J.}\ \bibnamefont
  {Brodsky}},\ }\href {\doibase 10.1016/0370-2693(79)90554-9} {\bibfield
  {journal} {\bibinfo  {journal} {Phys. Lett. B}\ }\textbf {\bibinfo {volume}
  {87}},\ \bibinfo {pages} {359} (\bibinfo {year} {1979})}\BibitemShut
  {NoStop}%
\bibitem [{\citenamefont {Efremov}\ and\ \citenamefont
  {Radyushkin}(1980)}]{Efremov:1979qk}%
  \BibitemOpen
  \bibfield  {author} {\bibinfo {author} {\bibfnamefont {A.~V.}\ \bibnamefont
  {Efremov}}\ and\ \bibinfo {author} {\bibfnamefont {A.~V.}\ \bibnamefont
  {Radyushkin}},\ }\href {\doibase 10.1016/0370-2693(80)90869-2} {\bibfield
  {journal} {\bibinfo  {journal} {Phys. Lett. B}\ }\textbf {\bibinfo {volume}
  {94}},\ \bibinfo {pages} {245} (\bibinfo {year} {1980})}\BibitemShut
  {NoStop}%
\bibitem [{\citenamefont {Lepage}\ and\ \citenamefont
  {Brodsky}(1980)}]{Lepage:1980fj}%
  \BibitemOpen
  \bibfield  {author} {\bibinfo {author} {\bibfnamefont {G.~P.}\ \bibnamefont
  {Lepage}}\ and\ \bibinfo {author} {\bibfnamefont {S.~J.}\ \bibnamefont
  {Brodsky}},\ }\href {\doibase 10.1103/PhysRevD.22.2157} {\bibfield  {journal}
  {\bibinfo  {journal} {Phys. Rev. D}\ }\textbf {\bibinfo {volume} {22}},\
  \bibinfo {pages} {2157} (\bibinfo {year} {1980})}\BibitemShut {NoStop}%
\bibitem [{\citenamefont {Descotes-Genon}\ and\ \citenamefont
  {Sachrajda}(2003{\natexlab{a}})}]{Descotes-Genon:2002crx}%
  \BibitemOpen
  \bibfield  {author} {\bibinfo {author} {\bibfnamefont {S.}~\bibnamefont
  {Descotes-Genon}}\ and\ \bibinfo {author} {\bibfnamefont {C.~T.}\
  \bibnamefont {Sachrajda}},\ }\href {\doibase 10.1016/S0550-3213(02)01066-0}
  {\bibfield  {journal} {\bibinfo  {journal} {Nucl. Phys. B}\ }\textbf
  {\bibinfo {volume} {650}},\ \bibinfo {pages} {356} (\bibinfo {year}
  {2003}{\natexlab{a}})},\ \Eprint {http://arxiv.org/abs/hep-ph/0209216}
  {arXiv:hep-ph/0209216} \BibitemShut {NoStop}%
\bibitem [{\citenamefont {Descotes-Genon}\ and\ \citenamefont
  {Sachrajda}(2003{\natexlab{b}})}]{Descotes-Genon:2002lal}%
  \BibitemOpen
  \bibfield  {author} {\bibinfo {author} {\bibfnamefont {S.}~\bibnamefont
  {Descotes-Genon}}\ and\ \bibinfo {author} {\bibfnamefont {C.~T.}\
  \bibnamefont {Sachrajda}},\ }\href {\doibase 10.1016/S0370-2693(03)00173-4}
  {\bibfield  {journal} {\bibinfo  {journal} {Phys. Lett. B}\ }\textbf
  {\bibinfo {volume} {557}},\ \bibinfo {pages} {213} (\bibinfo {year}
  {2003}{\natexlab{b}})},\ \Eprint {http://arxiv.org/abs/hep-ph/0212162}
  {arXiv:hep-ph/0212162} \BibitemShut {NoStop}%
\bibitem [{\citenamefont {Bosch}\ \emph {et~al.}(2003)\citenamefont {Bosch},
  \citenamefont {Hill}, \citenamefont {Lange},\ and\ \citenamefont
  {Neubert}}]{Bosch:2003fc}%
  \BibitemOpen
  \bibfield  {author} {\bibinfo {author} {\bibfnamefont {S.~W.}\ \bibnamefont
  {Bosch}}, \bibinfo {author} {\bibfnamefont {R.~J.}\ \bibnamefont {Hill}},
  \bibinfo {author} {\bibfnamefont {B.~O.}\ \bibnamefont {Lange}}, \ and\
  \bibinfo {author} {\bibfnamefont {M.}~\bibnamefont {Neubert}},\ }\href
  {\doibase 10.1103/PhysRevD.67.094014} {\bibfield  {journal} {\bibinfo
  {journal} {Phys. Rev. D}\ }\textbf {\bibinfo {volume} {67}},\ \bibinfo
  {pages} {094014} (\bibinfo {year} {2003})},\ \Eprint
  {http://arxiv.org/abs/hep-ph/0301123} {arXiv:hep-ph/0301123} \BibitemShut
  {NoStop}%
\bibitem [{\citenamefont {Beneke}\ \emph {et~al.}(2020)\citenamefont {Beneke},
  \citenamefont {Bobeth},\ and\ \citenamefont {Wang}}]{Beneke:2020fot}%
  \BibitemOpen
  \bibfield  {author} {\bibinfo {author} {\bibfnamefont {M.}~\bibnamefont
  {Beneke}}, \bibinfo {author} {\bibfnamefont {C.}~\bibnamefont {Bobeth}}, \
  and\ \bibinfo {author} {\bibfnamefont {Y.-M.}\ \bibnamefont {Wang}},\ }\href
  {\doibase 10.1007/JHEP12(2020)148} {\bibfield  {journal} {\bibinfo  {journal}
  {JHEP}\ }\textbf {\bibinfo {volume} {12}},\ \bibinfo {pages} {148} (\bibinfo
  {year} {2020})},\ \Eprint {http://arxiv.org/abs/2008.12494} {arXiv:2008.12494
  [hep-ph]} \BibitemShut {NoStop}%
\bibitem [{\citenamefont {Bosch}\ and\ \citenamefont
  {Buchalla}(2002)}]{Bosch:2002bv}%
  \BibitemOpen
  \bibfield  {author} {\bibinfo {author} {\bibfnamefont {S.~W.}\ \bibnamefont
  {Bosch}}\ and\ \bibinfo {author} {\bibfnamefont {G.}~\bibnamefont
  {Buchalla}},\ }\href {\doibase 10.1088/1126-6708/2002/08/054} {\bibfield
  {journal} {\bibinfo  {journal} {JHEP}\ }\textbf {\bibinfo {volume} {08}},\
  \bibinfo {pages} {054} (\bibinfo {year} {2002})},\ \Eprint
  {http://arxiv.org/abs/hep-ph/0208202} {arXiv:hep-ph/0208202} \BibitemShut
  {NoStop}%
\bibitem [{\citenamefont {Shen}\ \emph {et~al.}(2020)\citenamefont {Shen},
  \citenamefont {Wang},\ and\ \citenamefont {Wei}}]{Shen:2020hfq}%
  \BibitemOpen
  \bibfield  {author} {\bibinfo {author} {\bibfnamefont {Y.-L.}\ \bibnamefont
  {Shen}}, \bibinfo {author} {\bibfnamefont {Y.-M.}\ \bibnamefont {Wang}}, \
  and\ \bibinfo {author} {\bibfnamefont {Y.-B.}\ \bibnamefont {Wei}},\ }\href
  {\doibase 10.1007/JHEP12(2020)169} {\bibfield  {journal} {\bibinfo  {journal}
  {JHEP}\ }\textbf {\bibinfo {volume} {12}},\ \bibinfo {pages} {169} (\bibinfo
  {year} {2020})},\ \Eprint {http://arxiv.org/abs/2009.02723} {arXiv:2009.02723
  [hep-ph]} \BibitemShut {NoStop}%
\bibitem [{\citenamefont {Qin}\ \emph {et~al.}(2023)\citenamefont {Qin},
  \citenamefont {Shen}, \citenamefont {Wang},\ and\ \citenamefont
  {Wang}}]{Qin:2022rlk}%
  \BibitemOpen
  \bibfield  {author} {\bibinfo {author} {\bibfnamefont {Q.}~\bibnamefont
  {Qin}}, \bibinfo {author} {\bibfnamefont {Y.-L.}\ \bibnamefont {Shen}},
  \bibinfo {author} {\bibfnamefont {C.}~\bibnamefont {Wang}}, \ and\ \bibinfo
  {author} {\bibfnamefont {Y.-M.}\ \bibnamefont {Wang}},\ }\href {\doibase
  10.1103/PhysRevLett.131.091902} {\bibfield  {journal} {\bibinfo  {journal}
  {Phys. Rev. Lett.}\ }\textbf {\bibinfo {volume} {131}},\ \bibinfo {pages}
  {091902} (\bibinfo {year} {2023})},\ \Eprint
  {http://arxiv.org/abs/2207.02691} {arXiv:2207.02691 [hep-ph]} \BibitemShut
  {NoStop}%
\bibitem [{\citenamefont {Beneke}\ and\ \citenamefont
  {Feldmann}(2004)}]{Beneke:2003pa}%
  \BibitemOpen
  \bibfield  {author} {\bibinfo {author} {\bibfnamefont {M.}~\bibnamefont
  {Beneke}}\ and\ \bibinfo {author} {\bibfnamefont {T.}~\bibnamefont
  {Feldmann}},\ }\href {\doibase 10.1016/j.nuclphysb.2004.02.033} {\bibfield
  {journal} {\bibinfo  {journal} {Nucl. Phys. B}\ }\textbf {\bibinfo {volume}
  {685}},\ \bibinfo {pages} {249} (\bibinfo {year} {2004})},\ \Eprint
  {http://arxiv.org/abs/hep-ph/0311335} {arXiv:hep-ph/0311335} \BibitemShut
  {NoStop}%
\bibitem [{\citenamefont {Beneke}\ and\ \citenamefont
  {Jager}(2006)}]{Beneke:2005vv}%
  \BibitemOpen
  \bibfield  {author} {\bibinfo {author} {\bibfnamefont {M.}~\bibnamefont
  {Beneke}}\ and\ \bibinfo {author} {\bibfnamefont {S.}~\bibnamefont {Jager}},\
  }\href {\doibase 10.1016/j.nuclphysb.2006.06.010} {\bibfield  {journal}
  {\bibinfo  {journal} {Nucl. Phys. B}\ }\textbf {\bibinfo {volume} {751}},\
  \bibinfo {pages} {160} (\bibinfo {year} {2006})},\ \Eprint
  {http://arxiv.org/abs/hep-ph/0512351} {arXiv:hep-ph/0512351} \BibitemShut
  {NoStop}%
\bibitem [{\citenamefont {Beneke}\ and\ \citenamefont
  {Jager}(2007)}]{Beneke:2006mk}%
  \BibitemOpen
  \bibfield  {author} {\bibinfo {author} {\bibfnamefont {M.}~\bibnamefont
  {Beneke}}\ and\ \bibinfo {author} {\bibfnamefont {S.}~\bibnamefont {Jager}},\
  }\href {\doibase 10.1016/j.nuclphysb.2007.01.016} {\bibfield  {journal}
  {\bibinfo  {journal} {Nucl. Phys. B}\ }\textbf {\bibinfo {volume} {768}},\
  \bibinfo {pages} {51} (\bibinfo {year} {2007})},\ \Eprint
  {http://arxiv.org/abs/hep-ph/0610322} {arXiv:hep-ph/0610322} \BibitemShut
  {NoStop}%
\bibitem [{\citenamefont {Beneke}\ \emph {et~al.}(2010)\citenamefont {Beneke},
  \citenamefont {Huber},\ and\ \citenamefont {Li}}]{Beneke:2009ek}%
  \BibitemOpen
  \bibfield  {author} {\bibinfo {author} {\bibfnamefont {M.}~\bibnamefont
  {Beneke}}, \bibinfo {author} {\bibfnamefont {T.}~\bibnamefont {Huber}}, \
  and\ \bibinfo {author} {\bibfnamefont {X.-Q.}\ \bibnamefont {Li}},\ }\href
  {\doibase 10.1016/j.nuclphysb.2010.02.002} {\bibfield  {journal} {\bibinfo
  {journal} {Nucl. Phys. B}\ }\textbf {\bibinfo {volume} {832}},\ \bibinfo
  {pages} {109} (\bibinfo {year} {2010})},\ \Eprint
  {http://arxiv.org/abs/0911.3655} {arXiv:0911.3655 [hep-ph]} \BibitemShut
  {NoStop}%
\bibitem [{\citenamefont {Bell}\ \emph {et~al.}(2020)\citenamefont {Bell},
  \citenamefont {Beneke}, \citenamefont {Huber},\ and\ \citenamefont
  {Li}}]{Bell:2020qus}%
  \BibitemOpen
  \bibfield  {author} {\bibinfo {author} {\bibfnamefont {G.}~\bibnamefont
  {Bell}}, \bibinfo {author} {\bibfnamefont {M.}~\bibnamefont {Beneke}},
  \bibinfo {author} {\bibfnamefont {T.}~\bibnamefont {Huber}}, \ and\ \bibinfo
  {author} {\bibfnamefont {X.-Q.}\ \bibnamefont {Li}},\ }\href {\doibase
  10.1007/JHEP04(2020)055} {\bibfield  {journal} {\bibinfo  {journal} {JHEP}\
  }\textbf {\bibinfo {volume} {04}},\ \bibinfo {pages} {055} (\bibinfo {year}
  {2020})},\ \Eprint {http://arxiv.org/abs/2002.03262} {arXiv:2002.03262
  [hep-ph]} \BibitemShut {NoStop}%
\bibitem [{\citenamefont {Chay}\ and\ \citenamefont {Kim}(2004)}]{Chay:2003ju}%
  \BibitemOpen
  \bibfield  {author} {\bibinfo {author} {\bibfnamefont {J.}~\bibnamefont
  {Chay}}\ and\ \bibinfo {author} {\bibfnamefont {C.}~\bibnamefont {Kim}},\
  }\href {\doibase 10.1016/j.nuclphysb.2003.12.027} {\bibfield  {journal}
  {\bibinfo  {journal} {Nucl. Phys. B}\ }\textbf {\bibinfo {volume} {680}},\
  \bibinfo {pages} {302} (\bibinfo {year} {2004})},\ \Eprint
  {http://arxiv.org/abs/hep-ph/0301262} {arXiv:hep-ph/0301262} \BibitemShut
  {NoStop}%
\bibitem [{\citenamefont {Bauer}\ \emph {et~al.}(2004)\citenamefont {Bauer},
  \citenamefont {Pirjol}, \citenamefont {Rothstein},\ and\ \citenamefont
  {Stewart}}]{Bauer:2004tj}%
  \BibitemOpen
  \bibfield  {author} {\bibinfo {author} {\bibfnamefont {C.~W.}\ \bibnamefont
  {Bauer}}, \bibinfo {author} {\bibfnamefont {D.}~\bibnamefont {Pirjol}},
  \bibinfo {author} {\bibfnamefont {I.~Z.}\ \bibnamefont {Rothstein}}, \ and\
  \bibinfo {author} {\bibfnamefont {I.~W.}\ \bibnamefont {Stewart}},\ }\href
  {\doibase 10.1103/PhysRevD.70.054015} {\bibfield  {journal} {\bibinfo
  {journal} {Phys. Rev. D}\ }\textbf {\bibinfo {volume} {70}},\ \bibinfo
  {pages} {054015} (\bibinfo {year} {2004})},\ \Eprint
  {http://arxiv.org/abs/hep-ph/0401188} {arXiv:hep-ph/0401188} \BibitemShut
  {NoStop}%
\bibitem [{\citenamefont {Bauer}\ \emph {et~al.}(2002)\citenamefont {Bauer},
  \citenamefont {Pirjol},\ and\ \citenamefont {Stewart}}]{Bauer:2001yt}%
  \BibitemOpen
  \bibfield  {author} {\bibinfo {author} {\bibfnamefont {C.~W.}\ \bibnamefont
  {Bauer}}, \bibinfo {author} {\bibfnamefont {D.}~\bibnamefont {Pirjol}}, \
  and\ \bibinfo {author} {\bibfnamefont {I.~W.}\ \bibnamefont {Stewart}},\
  }\href {\doibase 10.1103/PhysRevD.65.054022} {\bibfield  {journal} {\bibinfo
  {journal} {Phys. Rev. D}\ }\textbf {\bibinfo {volume} {65}},\ \bibinfo
  {pages} {054022} (\bibinfo {year} {2002})},\ \Eprint
  {http://arxiv.org/abs/hep-ph/0109045} {arXiv:hep-ph/0109045} \BibitemShut
  {NoStop}%
\bibitem [{\citenamefont {Becher}\ \emph {et~al.}(2015)\citenamefont {Becher},
  \citenamefont {Broggio},\ and\ \citenamefont {Ferroglia}}]{Becher:2014oda}%
  \BibitemOpen
  \bibfield  {author} {\bibinfo {author} {\bibfnamefont {T.}~\bibnamefont
  {Becher}}, \bibinfo {author} {\bibfnamefont {A.}~\bibnamefont {Broggio}}, \
  and\ \bibinfo {author} {\bibfnamefont {A.}~\bibnamefont {Ferroglia}},\ }\href
  {\doibase 10.1007/978-3-319-14848-9} {\emph {\bibinfo {title} {{Introduction
  to Soft-Collinear Effective Theory}}}},\ Vol.\ \bibinfo {volume} {896}\
  (\bibinfo  {publisher} {Springer},\ \bibinfo {year} {2015})\ \Eprint
  {http://arxiv.org/abs/1410.1892} {arXiv:1410.1892 [hep-ph]} \BibitemShut
  {NoStop}%
\bibitem [{\citenamefont {Beneke}(2015)}]{Beneke:2015wfa}%
  \BibitemOpen
  \bibfield  {author} {\bibinfo {author} {\bibfnamefont {M.}~\bibnamefont
  {Beneke}},\ }\href {\doibase 10.1016/j.nuclphysbps.2015.03.021} {\bibfield
  {journal} {\bibinfo  {journal} {Nucl. Part. Phys. Proc.}\ }\textbf {\bibinfo
  {volume} {261-262}},\ \bibinfo {pages} {311} (\bibinfo {year} {2015})},\
  \Eprint {http://arxiv.org/abs/1501.07374} {arXiv:1501.07374 [hep-ph]}
  \BibitemShut {NoStop}%
\bibitem [{\citenamefont {Beneke}\ and\ \citenamefont
  {Feldmann}(2003)}]{Beneke:2002ni}%
  \BibitemOpen
  \bibfield  {author} {\bibinfo {author} {\bibfnamefont {M.}~\bibnamefont
  {Beneke}}\ and\ \bibinfo {author} {\bibfnamefont {T.}~\bibnamefont
  {Feldmann}},\ }\href {\doibase 10.1016/S0370-2693(02)03204-5} {\bibfield
  {journal} {\bibinfo  {journal} {Phys. Lett. B}\ }\textbf {\bibinfo {volume}
  {553}},\ \bibinfo {pages} {267} (\bibinfo {year} {2003})},\ \Eprint
  {http://arxiv.org/abs/hep-ph/0211358} {arXiv:hep-ph/0211358} \BibitemShut
  {NoStop}%
\bibitem [{\citenamefont {Wang}\ \emph {et~al.}(2022)\citenamefont {Wang},
  \citenamefont {Wang},\ and\ \citenamefont {Wei}}]{Wang:2021yrr}%
  \BibitemOpen
  \bibfield  {author} {\bibinfo {author} {\bibfnamefont {C.}~\bibnamefont
  {Wang}}, \bibinfo {author} {\bibfnamefont {Y.-M.}\ \bibnamefont {Wang}}, \
  and\ \bibinfo {author} {\bibfnamefont {Y.-B.}\ \bibnamefont {Wei}},\ }\href
  {\doibase 10.1007/JHEP02(2022)141} {\bibfield  {journal} {\bibinfo  {journal}
  {JHEP}\ }\textbf {\bibinfo {volume} {02}},\ \bibinfo {pages} {141} (\bibinfo
  {year} {2022})},\ \Eprint {http://arxiv.org/abs/2111.11811} {arXiv:2111.11811
  [hep-ph]} \BibitemShut {NoStop}%
\bibitem [{\citenamefont {Beneke}\ \emph
  {et~al.}(2018{\natexlab{a}})\citenamefont {Beneke}, \citenamefont {Garny},
  \citenamefont {Szafron},\ and\ \citenamefont {Wang}}]{Beneke:2018rbh}%
  \BibitemOpen
  \bibfield  {author} {\bibinfo {author} {\bibfnamefont {M.}~\bibnamefont
  {Beneke}}, \bibinfo {author} {\bibfnamefont {M.}~\bibnamefont {Garny}},
  \bibinfo {author} {\bibfnamefont {R.}~\bibnamefont {Szafron}}, \ and\
  \bibinfo {author} {\bibfnamefont {J.}~\bibnamefont {Wang}},\ }\href {\doibase
  10.1007/JHEP11(2018)112} {\bibfield  {journal} {\bibinfo  {journal} {JHEP}\
  }\textbf {\bibinfo {volume} {11}},\ \bibinfo {pages} {112} (\bibinfo {year}
  {2018}{\natexlab{a}})},\ \Eprint {http://arxiv.org/abs/1808.04742}
  {arXiv:1808.04742 [hep-ph]} \BibitemShut {NoStop}%
\bibitem [{\citenamefont {Lange}\ and\ \citenamefont
  {Neubert}(2003)}]{Lange:2003ff}%
  \BibitemOpen
  \bibfield  {author} {\bibinfo {author} {\bibfnamefont {B.~O.}\ \bibnamefont
  {Lange}}\ and\ \bibinfo {author} {\bibfnamefont {M.}~\bibnamefont
  {Neubert}},\ }\href {\doibase 10.1103/PhysRevLett.91.102001} {\bibfield
  {journal} {\bibinfo  {journal} {Phys. Rev. Lett.}\ }\textbf {\bibinfo
  {volume} {91}},\ \bibinfo {pages} {102001} (\bibinfo {year} {2003})},\
  \Eprint {http://arxiv.org/abs/hep-ph/0303082} {arXiv:hep-ph/0303082}
  \BibitemShut {NoStop}%
\bibitem [{\citenamefont {Braun}\ \emph {et~al.}(2004)\citenamefont {Braun},
  \citenamefont {Ivanov},\ and\ \citenamefont {Korchemsky}}]{Braun:2003wx}%
  \BibitemOpen
  \bibfield  {author} {\bibinfo {author} {\bibfnamefont {V.~M.}\ \bibnamefont
  {Braun}}, \bibinfo {author} {\bibfnamefont {D.~Y.}\ \bibnamefont {Ivanov}}, \
  and\ \bibinfo {author} {\bibfnamefont {G.~P.}\ \bibnamefont {Korchemsky}},\
  }\href {\doibase 10.1103/PhysRevD.69.034014} {\bibfield  {journal} {\bibinfo
  {journal} {Phys. Rev. D}\ }\textbf {\bibinfo {volume} {69}},\ \bibinfo
  {pages} {034014} (\bibinfo {year} {2004})},\ \Eprint
  {http://arxiv.org/abs/hep-ph/0309330} {arXiv:hep-ph/0309330} \BibitemShut
  {NoStop}%
\bibitem [{\citenamefont {Bell}\ \emph {et~al.}(2013)\citenamefont {Bell},
  \citenamefont {Feldmann}, \citenamefont {Wang},\ and\ \citenamefont
  {Yip}}]{Bell:2013tfa}%
  \BibitemOpen
  \bibfield  {author} {\bibinfo {author} {\bibfnamefont {G.}~\bibnamefont
  {Bell}}, \bibinfo {author} {\bibfnamefont {T.}~\bibnamefont {Feldmann}},
  \bibinfo {author} {\bibfnamefont {Y.-M.}\ \bibnamefont {Wang}}, \ and\
  \bibinfo {author} {\bibfnamefont {M.~W.~Y.}\ \bibnamefont {Yip}},\ }\href
  {\doibase 10.1007/JHEP11(2013)191} {\bibfield  {journal} {\bibinfo  {journal}
  {JHEP}\ }\textbf {\bibinfo {volume} {11}},\ \bibinfo {pages} {191} (\bibinfo
  {year} {2013})},\ \Eprint {http://arxiv.org/abs/1308.6114} {arXiv:1308.6114
  [hep-ph]} \BibitemShut {NoStop}%
\bibitem [{\citenamefont {Lee}\ and\ \citenamefont
  {Neubert}(2005)}]{Lee:2005gza}%
  \BibitemOpen
  \bibfield  {author} {\bibinfo {author} {\bibfnamefont {S.~J.}\ \bibnamefont
  {Lee}}\ and\ \bibinfo {author} {\bibfnamefont {M.}~\bibnamefont {Neubert}},\
  }\href {\doibase 10.1103/PhysRevD.72.094028} {\bibfield  {journal} {\bibinfo
  {journal} {Phys. Rev. D}\ }\textbf {\bibinfo {volume} {72}},\ \bibinfo
  {pages} {094028} (\bibinfo {year} {2005})},\ \Eprint
  {http://arxiv.org/abs/hep-ph/0509350} {arXiv:hep-ph/0509350} \BibitemShut
  {NoStop}%
\bibitem [{\citenamefont {Feldmann}\ \emph {et~al.}(2014)\citenamefont
  {Feldmann}, \citenamefont {Lange},\ and\ \citenamefont
  {Wang}}]{Feldmann:2014ika}%
  \BibitemOpen
  \bibfield  {author} {\bibinfo {author} {\bibfnamefont {T.}~\bibnamefont
  {Feldmann}}, \bibinfo {author} {\bibfnamefont {B.~O.}\ \bibnamefont {Lange}},
  \ and\ \bibinfo {author} {\bibfnamefont {Y.-M.}\ \bibnamefont {Wang}},\
  }\href {\doibase 10.1103/PhysRevD.89.114001} {\bibfield  {journal} {\bibinfo
  {journal} {Phys. Rev. D}\ }\textbf {\bibinfo {volume} {89}},\ \bibinfo
  {pages} {114001} (\bibinfo {year} {2014})},\ \Eprint
  {http://arxiv.org/abs/1404.1343} {arXiv:1404.1343 [hep-ph]} \BibitemShut
  {NoStop}%
\bibitem [{\citenamefont {Braun}\ \emph {et~al.}(2017)\citenamefont {Braun},
  \citenamefont {Ji},\ and\ \citenamefont {Manashov}}]{Braun:2017liq}%
  \BibitemOpen
  \bibfield  {author} {\bibinfo {author} {\bibfnamefont {V.~M.}\ \bibnamefont
  {Braun}}, \bibinfo {author} {\bibfnamefont {Y.}~\bibnamefont {Ji}}, \ and\
  \bibinfo {author} {\bibfnamefont {A.~N.}\ \bibnamefont {Manashov}},\ }\href
  {\doibase 10.1007/JHEP05(2017)022} {\bibfield  {journal} {\bibinfo  {journal}
  {JHEP}\ }\textbf {\bibinfo {volume} {05}},\ \bibinfo {pages} {022} (\bibinfo
  {year} {2017})},\ \Eprint {http://arxiv.org/abs/1703.02446} {arXiv:1703.02446
  [hep-ph]} \BibitemShut {NoStop}%
\bibitem [{\citenamefont {Feldmann}\ \emph {et~al.}(2023)\citenamefont
  {Feldmann}, \citenamefont {L\"ughausen},\ and\ \citenamefont
  {Seitz}}]{Feldmann:2023aml}%
  \BibitemOpen
  \bibfield  {author} {\bibinfo {author} {\bibfnamefont {T.}~\bibnamefont
  {Feldmann}}, \bibinfo {author} {\bibfnamefont {P.}~\bibnamefont
  {L\"ughausen}}, \ and\ \bibinfo {author} {\bibfnamefont {N.}~\bibnamefont
  {Seitz}},\ }\href {\doibase 10.1007/JHEP08(2023)075} {\bibfield  {journal}
  {\bibinfo  {journal} {JHEP}\ }\textbf {\bibinfo {volume} {08}},\ \bibinfo
  {pages} {075} (\bibinfo {year} {2023})},\ \Eprint
  {http://arxiv.org/abs/2306.14686} {arXiv:2306.14686 [hep-ph]} \BibitemShut
  {NoStop}%
\bibitem [{\citenamefont {Chetyrkin}\ \emph {et~al.}(1997)\citenamefont
  {Chetyrkin}, \citenamefont {Misiak},\ and\ \citenamefont
  {Munz}}]{Chetyrkin:1996vx}%
  \BibitemOpen
  \bibfield  {author} {\bibinfo {author} {\bibfnamefont {K.~G.}\ \bibnamefont
  {Chetyrkin}}, \bibinfo {author} {\bibfnamefont {M.}~\bibnamefont {Misiak}}, \
  and\ \bibinfo {author} {\bibfnamefont {M.}~\bibnamefont {Munz}},\ }\href
  {\doibase 10.1016/S0370-2693(97)00324-9} {\bibfield  {journal} {\bibinfo
  {journal} {Phys. Lett. B}\ }\textbf {\bibinfo {volume} {400}},\ \bibinfo
  {pages} {206} (\bibinfo {year} {1997})},\ \bibinfo {note} {[Erratum:
  Phys.Lett.B 425, 414 (1998)]},\ \Eprint {http://arxiv.org/abs/hep-ph/9612313}
  {arXiv:hep-ph/9612313} \BibitemShut {NoStop}%
\bibitem [{\citenamefont {Bobeth}\ \emph {et~al.}(2000)\citenamefont {Bobeth},
  \citenamefont {Misiak},\ and\ \citenamefont {Urban}}]{Bobeth:1999mk}%
  \BibitemOpen
  \bibfield  {author} {\bibinfo {author} {\bibfnamefont {C.}~\bibnamefont
  {Bobeth}}, \bibinfo {author} {\bibfnamefont {M.}~\bibnamefont {Misiak}}, \
  and\ \bibinfo {author} {\bibfnamefont {J.}~\bibnamefont {Urban}},\ }\href
  {\doibase 10.1016/S0550-3213(00)00007-9} {\bibfield  {journal} {\bibinfo
  {journal} {Nucl. Phys. B}\ }\textbf {\bibinfo {volume} {574}},\ \bibinfo
  {pages} {291} (\bibinfo {year} {2000})},\ \Eprint
  {http://arxiv.org/abs/hep-ph/9910220} {arXiv:hep-ph/9910220} \BibitemShut
  {NoStop}%
\bibitem [{\citenamefont {Bobeth}\ \emph {et~al.}(2004)\citenamefont {Bobeth},
  \citenamefont {Gambino}, \citenamefont {Gorbahn},\ and\ \citenamefont
  {Haisch}}]{Bobeth:2003at}%
  \BibitemOpen
  \bibfield  {author} {\bibinfo {author} {\bibfnamefont {C.}~\bibnamefont
  {Bobeth}}, \bibinfo {author} {\bibfnamefont {P.}~\bibnamefont {Gambino}},
  \bibinfo {author} {\bibfnamefont {M.}~\bibnamefont {Gorbahn}}, \ and\
  \bibinfo {author} {\bibfnamefont {U.}~\bibnamefont {Haisch}},\ }\href
  {\doibase 10.1088/1126-6708/2004/04/071} {\bibfield  {journal} {\bibinfo
  {journal} {JHEP}\ }\textbf {\bibinfo {volume} {04}},\ \bibinfo {pages} {071}
  (\bibinfo {year} {2004})},\ \Eprint {http://arxiv.org/abs/hep-ph/0312090}
  {arXiv:hep-ph/0312090} \BibitemShut {NoStop}%
\bibitem [{\citenamefont {Huber}\ \emph {et~al.}(2006)\citenamefont {Huber},
  \citenamefont {Lunghi}, \citenamefont {Misiak},\ and\ \citenamefont
  {Wyler}}]{Huber:2005ig}%
  \BibitemOpen
  \bibfield  {author} {\bibinfo {author} {\bibfnamefont {T.}~\bibnamefont
  {Huber}}, \bibinfo {author} {\bibfnamefont {E.}~\bibnamefont {Lunghi}},
  \bibinfo {author} {\bibfnamefont {M.}~\bibnamefont {Misiak}}, \ and\ \bibinfo
  {author} {\bibfnamefont {D.}~\bibnamefont {Wyler}},\ }\href {\doibase
  10.1016/j.nuclphysb.2006.01.037} {\bibfield  {journal} {\bibinfo  {journal}
  {Nucl. Phys. B}\ }\textbf {\bibinfo {volume} {740}},\ \bibinfo {pages} {105}
  (\bibinfo {year} {2006})},\ \Eprint {http://arxiv.org/abs/hep-ph/0512066}
  {arXiv:hep-ph/0512066} \BibitemShut {NoStop}%
\bibitem [{\citenamefont {Beneke}\ \emph
  {et~al.}(2018{\natexlab{b}})\citenamefont {Beneke}, \citenamefont {Braun},
  \citenamefont {Ji},\ and\ \citenamefont {Wei}}]{Beneke:2018wjp}%
  \BibitemOpen
  \bibfield  {author} {\bibinfo {author} {\bibfnamefont {M.}~\bibnamefont
  {Beneke}}, \bibinfo {author} {\bibfnamefont {V.~M.}\ \bibnamefont {Braun}},
  \bibinfo {author} {\bibfnamefont {Y.}~\bibnamefont {Ji}}, \ and\ \bibinfo
  {author} {\bibfnamefont {Y.-B.}\ \bibnamefont {Wei}},\ }\href {\doibase
  10.1007/JHEP07(2018)154} {\bibfield  {journal} {\bibinfo  {journal} {JHEP}\
  }\textbf {\bibinfo {volume} {07}},\ \bibinfo {pages} {154} (\bibinfo {year}
  {2018}{\natexlab{b}})},\ \Eprint {http://arxiv.org/abs/1804.04962}
  {arXiv:1804.04962 [hep-ph]} \BibitemShut {NoStop}%
\bibitem [{\citenamefont {Galda}\ \emph {et~al.}(2022)\citenamefont {Galda},
  \citenamefont {Neubert},\ and\ \citenamefont {Wang}}]{Galda:2022dhp}%
  \BibitemOpen
  \bibfield  {author} {\bibinfo {author} {\bibfnamefont {A.~M.}\ \bibnamefont
  {Galda}}, \bibinfo {author} {\bibfnamefont {M.}~\bibnamefont {Neubert}}, \
  and\ \bibinfo {author} {\bibfnamefont {X.}~\bibnamefont {Wang}},\ }\href
  {\doibase 10.1007/JHEP07(2022)148} {\bibfield  {journal} {\bibinfo  {journal}
  {JHEP}\ }\textbf {\bibinfo {volume} {07}},\ \bibinfo {pages} {148} (\bibinfo
  {year} {2022})},\ \Eprint {http://arxiv.org/abs/2203.08202} {arXiv:2203.08202
  [hep-ph]} \BibitemShut {NoStop}%
\bibitem [{\citenamefont {Feldmann}\ \emph {et~al.}(2022)\citenamefont
  {Feldmann}, \citenamefont {L\"ughausen},\ and\ \citenamefont {van
  Dyk}}]{Feldmann:2022uok}%
  \BibitemOpen
  \bibfield  {author} {\bibinfo {author} {\bibfnamefont {T.}~\bibnamefont
  {Feldmann}}, \bibinfo {author} {\bibfnamefont {P.}~\bibnamefont
  {L\"ughausen}}, \ and\ \bibinfo {author} {\bibfnamefont {D.}~\bibnamefont
  {van Dyk}},\ }\href {\doibase 10.1007/JHEP10(2022)162} {\bibfield  {journal}
  {\bibinfo  {journal} {JHEP}\ }\textbf {\bibinfo {volume} {10}},\ \bibinfo
  {pages} {162} (\bibinfo {year} {2022})},\ \Eprint
  {http://arxiv.org/abs/2203.15679} {arXiv:2203.15679 [hep-ph]} \BibitemShut
  {NoStop}%
\bibitem [{\citenamefont {Braun}\ \emph {et~al.}(2018)\citenamefont {Braun},
  \citenamefont {Ji},\ and\ \citenamefont {Manashov}}]{Braun:2018fiz}%
  \BibitemOpen
  \bibfield  {author} {\bibinfo {author} {\bibfnamefont {V.~M.}\ \bibnamefont
  {Braun}}, \bibinfo {author} {\bibfnamefont {Y.}~\bibnamefont {Ji}}, \ and\
  \bibinfo {author} {\bibfnamefont {A.~N.}\ \bibnamefont {Manashov}},\ }\href
  {\doibase 10.1007/JHEP06(2018)017} {\bibfield  {journal} {\bibinfo  {journal}
  {JHEP}\ }\textbf {\bibinfo {volume} {06}},\ \bibinfo {pages} {017} (\bibinfo
  {year} {2018})},\ \Eprint {http://arxiv.org/abs/1804.06289} {arXiv:1804.06289
  [hep-th]} \BibitemShut {NoStop}%
\bibitem [{\citenamefont {Wang}\ \emph {et~al.}(2020)\citenamefont {Wang},
  \citenamefont {Wang}, \citenamefont {Xu},\ and\ \citenamefont
  {Zhao}}]{Wang:2019msf}%
  \BibitemOpen
  \bibfield  {author} {\bibinfo {author} {\bibfnamefont {W.}~\bibnamefont
  {Wang}}, \bibinfo {author} {\bibfnamefont {Y.-M.}\ \bibnamefont {Wang}},
  \bibinfo {author} {\bibfnamefont {J.}~\bibnamefont {Xu}}, \ and\ \bibinfo
  {author} {\bibfnamefont {S.}~\bibnamefont {Zhao}},\ }\href {\doibase
  10.1103/PhysRevD.102.011502} {\bibfield  {journal} {\bibinfo  {journal}
  {Phys. Rev. D}\ }\textbf {\bibinfo {volume} {102}},\ \bibinfo {pages}
  {011502} (\bibinfo {year} {2020})},\ \Eprint
  {http://arxiv.org/abs/1908.09933} {arXiv:1908.09933 [hep-ph]} \BibitemShut
  {NoStop}%
\bibitem [{\citenamefont {Aoki}\ \emph {et~al.}(2022)\citenamefont {Aoki} \emph
  {et~al.}}]{FlavourLatticeAveragingGroupFLAG:2021npn}%
  \BibitemOpen
  \bibfield  {author} {\bibinfo {author} {\bibfnamefont {Y.}~\bibnamefont
  {Aoki}} \emph {et~al.} (\bibinfo {collaboration} {Flavour Lattice Averaging
  Group (FLAG)}),\ }\href {\doibase 10.1140/epjc/s10052-022-10536-1} {\bibfield
   {journal} {\bibinfo  {journal} {Eur. Phys. J. C}\ }\textbf {\bibinfo
  {volume} {82}},\ \bibinfo {pages} {869} (\bibinfo {year} {2022})},\ \Eprint
  {http://arxiv.org/abs/2111.09849} {arXiv:2111.09849 [hep-lat]} \BibitemShut
  {NoStop}%
\bibitem [{\citenamefont {Bali}\ \emph {et~al.}(2019)\citenamefont {Bali},
  \citenamefont {Braun}, \citenamefont {B\"urger}, \citenamefont {G\"ockeler},
  \citenamefont {Gruber}, \citenamefont {Hutzler}, \citenamefont {Korcyl},
  \citenamefont {Sch\"afer}, \citenamefont {Sternbeck},\ and\ \citenamefont
  {Wein}}]{RQCD:2019osh}%
  \BibitemOpen
  \bibfield  {author} {\bibinfo {author} {\bibfnamefont {G.~S.}\ \bibnamefont
  {Bali}}, \bibinfo {author} {\bibfnamefont {V.~M.}\ \bibnamefont {Braun}},
  \bibinfo {author} {\bibfnamefont {S.}~\bibnamefont {B\"urger}}, \bibinfo
  {author} {\bibfnamefont {M.}~\bibnamefont {G\"ockeler}}, \bibinfo {author}
  {\bibfnamefont {M.}~\bibnamefont {Gruber}}, \bibinfo {author} {\bibfnamefont
  {F.}~\bibnamefont {Hutzler}}, \bibinfo {author} {\bibfnamefont
  {P.}~\bibnamefont {Korcyl}}, \bibinfo {author} {\bibfnamefont
  {A.}~\bibnamefont {Sch\"afer}}, \bibinfo {author} {\bibfnamefont
  {A.}~\bibnamefont {Sternbeck}}, \ and\ \bibinfo {author} {\bibfnamefont
  {P.}~\bibnamefont {Wein}} (\bibinfo {collaboration} {RQCD}),\ }\href
  {\doibase 10.1007/JHEP08(2019)065} {\bibfield  {journal} {\bibinfo  {journal}
  {JHEP}\ }\textbf {\bibinfo {volume} {08}},\ \bibinfo {pages} {065} (\bibinfo
  {year} {2019})},\ \bibinfo {note} {[Addendum: JHEP 11, 037 (2020)]},\ \Eprint
  {http://arxiv.org/abs/1903.08038} {arXiv:1903.08038 [hep-lat]} \BibitemShut
  {NoStop}%
\bibitem [{\citenamefont {Mueller}(1994)}]{Mueller:1993hg}%
  \BibitemOpen
  \bibfield  {author} {\bibinfo {author} {\bibfnamefont {D.}~\bibnamefont
  {Mueller}},\ }\href {\doibase 10.1103/PhysRevD.49.2525} {\bibfield  {journal}
  {\bibinfo  {journal} {Phys. Rev. D}\ }\textbf {\bibinfo {volume} {49}},\
  \bibinfo {pages} {2525} (\bibinfo {year} {1994})}\BibitemShut {NoStop}%
\bibitem [{\citenamefont {Mueller}(1995)}]{Mueller:1994cn}%
  \BibitemOpen
  \bibfield  {author} {\bibinfo {author} {\bibfnamefont {D.}~\bibnamefont
  {Mueller}},\ }\href {\doibase 10.1103/PhysRevD.51.3855} {\bibfield  {journal}
  {\bibinfo  {journal} {Phys. Rev. D}\ }\textbf {\bibinfo {volume} {51}},\
  \bibinfo {pages} {3855} (\bibinfo {year} {1995})},\ \Eprint
  {http://arxiv.org/abs/hep-ph/9411338} {arXiv:hep-ph/9411338} \BibitemShut
  {NoStop}%
\bibitem [{\citenamefont {Agaev}\ \emph {et~al.}(2011)\citenamefont {Agaev},
  \citenamefont {Braun}, \citenamefont {Offen},\ and\ \citenamefont
  {Porkert}}]{Agaev:2010aq}%
  \BibitemOpen
  \bibfield  {author} {\bibinfo {author} {\bibfnamefont {S.~S.}\ \bibnamefont
  {Agaev}}, \bibinfo {author} {\bibfnamefont {V.~M.}\ \bibnamefont {Braun}},
  \bibinfo {author} {\bibfnamefont {N.}~\bibnamefont {Offen}}, \ and\ \bibinfo
  {author} {\bibfnamefont {F.~A.}\ \bibnamefont {Porkert}},\ }\href {\doibase
  10.1103/PhysRevD.83.054020} {\bibfield  {journal} {\bibinfo  {journal} {Phys.
  Rev. D}\ }\textbf {\bibinfo {volume} {83}},\ \bibinfo {pages} {054020}
  (\bibinfo {year} {2011})},\ \Eprint {http://arxiv.org/abs/1012.4671}
  {arXiv:1012.4671 [hep-ph]} \BibitemShut {NoStop}%
\bibitem [{\citenamefont {Wang}\ and\ \citenamefont
  {Shen}(2017)}]{Wang:2017ijn}%
  \BibitemOpen
  \bibfield  {author} {\bibinfo {author} {\bibfnamefont {Y.-M.}\ \bibnamefont
  {Wang}}\ and\ \bibinfo {author} {\bibfnamefont {Y.-L.}\ \bibnamefont
  {Shen}},\ }\href {\doibase 10.1007/JHEP12(2017)037} {\bibfield  {journal}
  {\bibinfo  {journal} {JHEP}\ }\textbf {\bibinfo {volume} {12}},\ \bibinfo
  {pages} {037} (\bibinfo {year} {2017})},\ \Eprint
  {http://arxiv.org/abs/1706.05680} {arXiv:1706.05680 [hep-ph]} \BibitemShut
  {NoStop}%
\bibitem [{\citenamefont {Aaij}\ \emph
  {et~al.}(2014{\natexlab{b}})\citenamefont {Aaij} \emph
  {et~al.}}]{LHCb:2014mit}%
  \BibitemOpen
  \bibfield  {author} {\bibinfo {author} {\bibfnamefont {R.}~\bibnamefont
  {Aaij}} \emph {et~al.} (\bibinfo {collaboration} {LHCb}),\ }\href {\doibase
  10.1007/JHEP09(2014)177} {\bibfield  {journal} {\bibinfo  {journal} {JHEP}\
  }\textbf {\bibinfo {volume} {09}},\ \bibinfo {pages} {177} (\bibinfo {year}
  {2014}{\natexlab{b}})},\ \Eprint {http://arxiv.org/abs/1408.0978}
  {arXiv:1408.0978 [hep-ex]} \BibitemShut {NoStop}%
\end{thebibliography}%

\end{document}